\documentclass[12pt]{article}
\usepackage{times}

\usepackage{epsf}
\def\beq{\begin{equation}}
\def\eeq{\end{equation}}
\def\bea{\begin{eqnarray}}
\def\eea{\end{eqnarray}}
\def\ve{\vert}
\def\vel{\left|}
\def\ver{\right|}
\def\nnb{\nonumber}
\def\ga{\left(}
\def\dr{\right)}

\def\rar{\rightarrow}
\def\nnb{\nonumber}
\def\la{\langle}
\def\ra{\rangle}
\def\ba{\begin{array}}
\def\ea{\end{array}}
\def\bea{\begin{eqnarray}}
\def\eea{\end{eqnarray}}

\title{ {\bf
$B\rightarrow \tau^+ \tau^- \gamma $  decay in the general two Higgs doublet 
model including the neutral Higgs boson effects }}
\author{\vspace{1cm}\\
         {\bf G. Erkol} 
         \thanks{E-mail address:
        gurerk@newton.physics.metu.edu.tr} \, \, and \, \, 
        {\bf G. Turan}
        \thanks{E-mail address:
        gsevgur@metu.edu.tr}\,\,}

\date{}

\begin{document}
\setlength{\baselineskip}{24pt}
\maketitle
\setlength{\baselineskip}{7mm}
\begin{abstract}
We investigate the differential branching ratio, branching ratio,
differential forward-backward asymmetry, the forward-backward
asymmetry of the lepton pair and the lepton polarization asymmetry of  the exclusive  
$B\rightarrow  \tau^+ \tau^- \gamma$ decay in the
general two  Higgs doublet model including the neutral Higgs boson effects.
We analyse the dependencies of these quantities  on the model parameters and
also on the neutral Higgs boson effects. 
We found that they get considerable
enhancement from the two Higgs doublet model compared to the standard model 
and neutral Higgs boson effects are quite sizable.
\end{abstract} 
\thispagestyle{empty}
\newpage
\setcounter{page}{1}
\section{Introduction}
Rare B-meson decays are induced by the flavor--changing neutral currents (FCNC)
and they occur only through electroweak loops in the standard model (SM). Thus,
on the one hand, they provide fertile testing ground for the SM and on the
other hand, they offer a complementary strategy for constraining new physics
beyond the SM, such as the two Higgs doublet model (2HDM), minimal
supersymmetric extension of the SM \cite{Hewett}, ect.  From the experimental 
point of view,
studying rare  $B$ meson decays can provide essential information
about the poorly known  parameters of the SM, like the elements of the 
Cabibbo--Kobayashi--Maskawa (CKM) matrix, the leptonic decay constants 
etc.

In this work, we study the radiative $B \rar \tau^+ \tau^-\gamma $ decay in
the general  two-Higgs doublet model (2HDM). 
It is induced by the pure-leptonic decay  $B \rar \tau^+ \tau^-$ and in principle,
the latter can be used to determine the decay constant $f_{B}$ \cite{Buchalla}, 
as well as
the fundamental parameters of the SM. However, it is well known that 
processes  $B \rar \ell^+ \ell^-$ are helicity suppressed for light lepton
modes, having branching ratios of the order of $10^{-9}$ for $\ell=\mu$
and $10^{-10}$ for $\ell=e$ channels \cite{Eilam1}. Although
the  $B \rar \tau^+ \tau^-$ channel is free from this suppression, 
its observation is difficult due to low efficiency. If a photon line is
attached to any of the charged lines (see Fig.1), the pure leptonic
processes $B \rar \ell^+ \ell^-$ change into the corresponding radiative ones,  
$B \rar \ell^+ \ell^- \gamma$, so helicity suppression is
overcome and larger branching ratios are expected. Depending on whether the 
photon is released from the initial quark
or final lepton lines, there exist two different types of contributions,
namely  the so-called "the structure dependent"
(SD) and the "internal  Bremsstrahlung" (IB)
respectively, while contributions coming from the release of the free 
photon from 
any charged internal line will be suppressed by a factor of $m^2_b/M^2_W$.  
The SD contribution is governed by the vector
and axial vector form factors and it is free from the helicity  suppression. 
Therefore, it could enhance the  decay rates of the radiative 
processes $B \rar \tau^+ \tau^- \gamma$  in
comparison to the decay  rates of the pure leptonic ones $B \rar \tau^+ \tau^-$.   
As for the IB part of the contribution, it
is proportional to the ratio  $m_{\ell}/m_{B}$ and therefore it is still
helicity suppressed  for the light charged lepton modes while it enhances the amplitude
considerably for $\ell=\tau$ mode. Indeed, $B \rar \ell^+
\ell^-\gamma $ decay have been investigated in the framework of the SM
for light and heavy  lepton modes \cite{Eilam1}-\cite{Aliev2}, 
as well as in the models beyond the SM   
\cite{Iltan1,Aliev3}, and it was found that in the  SM 
$BR(B \rar \ell^+ \ell^-\gamma)=2.35 \times
10^{-9} \cite{Eilam1}, 1.90 \times 10^{-9} \cite{Aliev1}, 9.54 \times10^{-9}$
\cite{Aliev2}, for $\ell=e,\mu , \tau$,
respectively. With long distance contributions, $BR(B \rar \tau^+
\tau^-\gamma)=1.52 \times 10^{-8}$ was obtained \cite{Aliev2}. 
In 2HDM, in contrast to the
channels with light leptons, the channel  $B \rar \tau^+ \tau^-\gamma$ receives 
additional contributions from the neutral Higgs boson (NHB) exchanges, in
addition to SD and IB ones. 
In \cite{Iltan1},  $B \rar \tau^+ \tau^-\gamma $ decay is investigated in 
model I and II types of the 2HDM including NHB effects and shown that these 
effects are sizable when $\tan \beta $ is large.
Our aim in this work is to study the sensitivity of the physically measurable
quantities, such as branching ratio, photon energy density, forward-backward
asymmetry of the final lepton and lepton polarization asymmetry, to the NHB effects, 
as well as to the model III parameters, like the Yukawa couplings  
$\bar{\xi}^{D}_{N,bb}$ and $\bar{\xi}^{D}_{N,\tau \tau}$. 
 
The work is organized as follows. In section 2, after a brief summary about
the main points of the general 2HDM, we first  present the leading order (LO)
QCD corrected effective Hamiltonian for the process $b \rar s \tau^+
\tau^-$, including the NHB effects and then give the matrix element for the
exclusive $B \rar \tau^+ \tau^- \gamma$  decay, together with the explicit
expressions for the  double differential decay width, photon energy
distribution, forward-backward asymmetry and the polarization asymmetry of the 
final lepton $\tau^{-}$. Section 3 is
devoted to the numerical analysis of the dependencies of these observables
on the model III parameters and also on  NHB effects. Finally, in the Appendix,
we give the explicit forms of the operators appearing in the Hamiltonian and
the corresponding Wilson coefficients. 
\section{The $B \rar \tau^+ \tau^- \gamma$ decay in the 
framework of the general 2HDM}
The 2HDM is the minimal extension of the SM, which consists of adding a second 
doublet to the Higgs sector. In this model,  there are one charged Higgs scalar, 
two neutral Higgs scalars and one neutral Higgs pseudoscalar.
The general Yukawa Lagrangian, which is responsible for the
interactions of quarks  with gauge bosons,  can be written as
\begin{eqnarray}
{\cal{L}}_{Y}&=&\eta^{U}_{ij} \bar{Q}_{i L} \tilde{\phi_{1}} U_{j R}+
\eta^{D}_{ij} \bar{Q}_{i L} \phi_{1} D_{j R}+
\xi^{U\, \dagger}_{ij} \bar{Q}_{i L} \tilde{\phi_{2}} U_{j R}+
\xi^{D}_{ij} \bar{Q}_{i L} \phi_{2} D_{j R}
+ h.c. \,\,\, ,
\label{lagrangian}
\end{eqnarray}
where $i,j$  are family indices of quarks , $L$ and $R$ 
denote chiral projections $L(R)=1/2(1\mp \gamma_5)$, $\phi_{m}$ for $m=1,2$, 
are the two scalar doublets, $Q_{i L}$ are quark  
doublets, $U_{j R}$, $D_{j R}$ are the corresponding quark 
singlets, $\eta^{U,D}_{ij}$ and $\xi^{U,D}_{ij}$ are the matrices 
of the Yukawa couplings. 
The Yukawa Lagrangian in Eq. (\ref{lagrangian}) opens up the possibility 
that there appear tree-level FCNC. In the SM and in  
model I and model II types of the 2HDM, such  FCNC at tree level are forbidden by the 
GIM mechanism \cite{GIM} and by  an {\it ad hoc} discrete symmetry
\cite{Glashow}, respectively. However, tree-level FCNC
are permitted in the general 2HDM, and this type of 2HDM is referred to as model III 
in the literature.

In this model, it is possible to choose $\phi_1$ and $\phi_2$ in the following form
\begin{eqnarray}
\phi_{1}=\frac{1}{\sqrt{2}}\left[\left(\begin{array}{c c} 
0\\v+H^{0}\end{array}\right)\; + \left(\begin{array}{c c} 
\sqrt{2} \chi^{+}\\ i \chi^{0}\end{array}\right) \right]\, ; 
\phi_{2}=\frac{1}{\sqrt{2}}\left(\begin{array}{c c} 
\sqrt{2} H^{+}\\ H_1+i H_2 \end{array}\right) \,\, 
\label{choice}
\end{eqnarray}
with the vacuum expectation values,  
\begin{eqnarray}
<\phi_{1}>=\frac{1}{\sqrt{2}}\left(\begin{array}{c c} 
0\\v\end{array}\right) \,  \, ; 
<\phi_{2}>=0 \,\, .
\label{choice2}
\end{eqnarray}
With this choice, the SM particles can be collected in the first doublet 
and the new particles in the second one. Further, we take $H_{1}$, $H_{2}$ 
as the mass eigenstates $h^{0}$, $A^{0}$ respectively. Note that, at tree 
level, there is no mixing among CP even neutral Higgs bosons, namely 
the SM one, $H^0$, and beyond, $h^{0}$.

The part which produces FCNC at tree level is  
\begin{eqnarray}
{\cal{L}}_{Y,FC}=
\xi^{U\,\dagger}_{ij} \bar{Q}_{i L} \tilde{\phi_{2}} U_{j R}+
\xi^{D}_{ij} \bar{Q}_{i L} \phi_{2} D_{j R} +
\xi^{D}_{kl} \bar{l}_{k L} \phi_{2} E_{l R} + h.c. \,\, .
\label{lagrangianFC}
\end{eqnarray}
In Eq.(\ref{lagrangianFC}), the couplings  $\xi^{U,D}$ for the
flavor-changing charged interactions are 
\begin{eqnarray}
\xi^{U}_{ch}&=& \xi_{neutral} \,\, V_{CKM} \nonumber \,\, ,\\
\xi^{D}_{ch}&=& V_{CKM} \,\, \xi_{neutral} \,\, ,
\label{ksi1} 
\end{eqnarray}
where  $\xi^{U,D}_{neutral}$ 
is defined by the expression
\begin{eqnarray}
\xi^{U (D)}_{N}=(V_{R(L)}^{U (D)})^{-1} \xi^{U,(D)} V_{L(R)}^{U (D)}\,, 
\label{ksineut}
\end{eqnarray}
and $\xi^{U,D}_{neutral}$ is denoted as $\xi^{U,D}_{N}$. 
Here the charged couplings are  the linear combinations of neutral 
couplings multiplied by $V_{CKM}$ matrix elements (see \cite{Aliev4} for
details). 

After this brief summary about the general 2HDM, now we would like to present the 
calculation of the matrix element for the $B \rar \tau^+ \tau^- \gamma$ decay .
For a general investigation of the  $B \rar \tau^+ \tau^- \gamma$ decay, 
we start with the  LO QCD corrected 
effective Hamiltonian which induces the corresponding quark level process
$b \rar s \,  \tau^+ \tau^-$, given by \cite{Dai}
\beq
{\cal H}=\frac{-4\, G_F}{ \sqrt{2}} V_{tb} V_{ts}^*
\Bigg{\{}\sum_{ i =1}^{10} C_{i}(\mu ) O_{i}(\mu)+
\sum_{ i =1}^{10} C_{Q_{i}}(\mu )Q_{i}(\mu) \Bigg{\}} \, ,
\label{H1}
\eeq
where $O_{i}$ are current-current $(i=1,2)$, penguin $(i=1,..,6)$, 
magnetic penguin $(i=7,8)$ and semileptonic $(i=9,10)$ operators . The 
additional operators $Q_{i}, (i=1,..,10)$ are due to the 
NHB exchange diagrams, which give considerable contributions in the case that the 
lepton pair is $\tau^{+}\tau^{-}$ \cite{Dai}. 
$C_{i}(\mu )$ and $C_{Q_{i}}(\mu )$ are Wilson coefficients renormalized at 
the scale $\mu$. All these operators and the Wilson coefficients, together 
with their initial values calculated at $\mu=m_W$ in the SM and also the 
additional coefficients coming from the new Higgs scalars are presented in 
Appendix A.

\begin{figure}[htb]
\vskip -5.0truein
\centering
\epsfxsize=6.8in
\leavevmode\epsffile{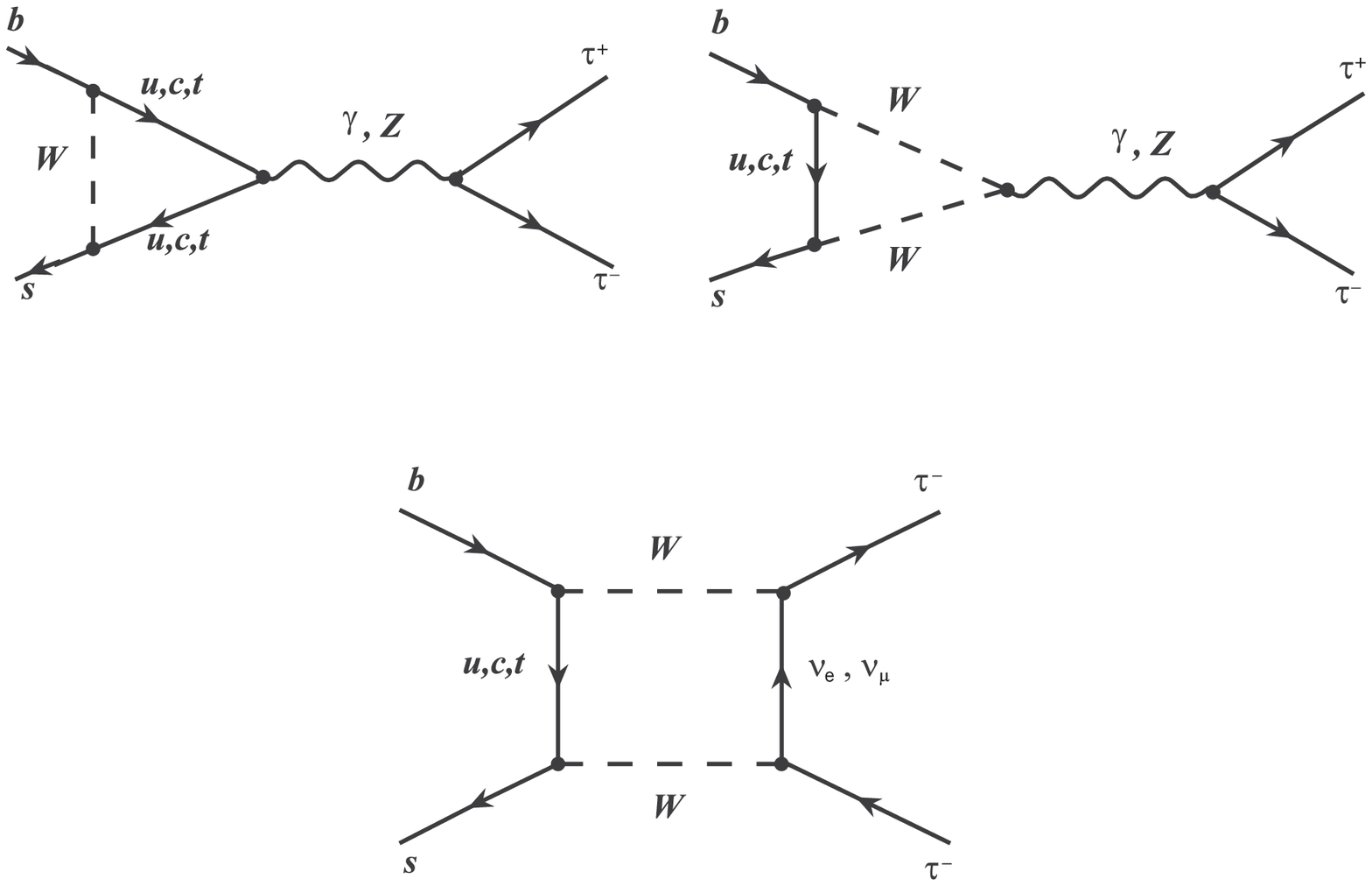}
\vskip -5.0truein
\caption[]{ Feynman diagrams  for $b \rar s \tau^+ \tau^-$ in the SM}
\label{feyndia}
\end{figure}

The short distance contributions for $B \rar \tau^+ \tau^- \gamma $ decay
come from the box, Z and photon penguin diagrams, which are obtained from 
the diagrams of Fig. (\ref{feyndia}) by attaching an additional photon line
either to the  initial quark lines that contribute to  the SD part of the 
amplitude,  or to the final lepton lines, which give the so-called IB
part of the amplitude. Following this framework,
the general form of the gauge invariant amplitude corresponding to
Fig.(\ref{feyndia}) can be written as the sum of the SD and IB parts
\bea
{\cal M} (B \rar \tau^+ \tau^- \gamma ) & = & {\cal M}_{SD}+ {\cal M}_{IB} \,,
\eea
where 
\bea
\label{Msd}
{\cal M}_{SD} &=& \frac{\alpha G_F}{2 \sqrt{2} \, \pi} V_{tb} V_{ts}^* 
\frac{e}{m_B^2} \,\Bigg\{
\bar \tau \gamma^\mu  \tau \, \Big[
A_1 \epsilon_{\mu \nu \alpha \beta} 
\varepsilon^{\ast\nu} q^\alpha k^\beta + 
i \, A_2 \Big( \varepsilon_\mu^\ast (k q) - 
(\varepsilon^\ast q ) k_\mu \Big) \Big] \nnb \\
&+& \bar \tau \gamma^\mu \gamma_5 \tau \, \Big[
B_1 \epsilon_{\mu \nu \alpha \beta} 
\varepsilon^{\ast\nu} q^\alpha k^\beta 
+ i \, B_2 \Big( \varepsilon_\mu^\ast (k q) - 
(\varepsilon^\ast q ) k_\mu \Big) \Big] \Bigg\} \, ,
\eea
and
\bea
\label{Mib}
{\cal M}_{IB} &=& \frac{\alpha G_F}{2 \sqrt{2} \, \pi} V_{tb} V_{ts}^*  
e f_B i \,\Bigg\{
F\, \bar \tau  \Bigg(
\frac{{\not\!\varepsilon}^\ast {\not\!p}_B}{2 p_1 k} - 
\frac{{\not\!p}_B {\not\!\varepsilon}^\ast}{2 p_2 k} \Bigg) 
\gamma_5 \tau \nnb \\
&+& F_1 \, \bar \tau \Bigg[
\frac{{\not\!\varepsilon}^\ast {\not\!p}_B}{2 p_1 k} -
\frac{{\not\!p}_B {\not\!\varepsilon}^\ast}{2 p_2 k} +
2 m_\tau \Bigg(\frac{1}{2 p_1 k} + \frac{1}{2 p_2 k}\Bigg)
{\not\!\varepsilon}^\ast \Bigg] \tau \Bigg\}~,
\eea
where
\bea
A_1 &=& \frac{-2}{q^2} m_b C^{eff}_7 g_1 + C^{eff}_9 g ~, \nnb \\
A_2 &=& \frac{-2}{q^2} m_b C^{eff}_7 f_1 + C^{eff}_9 f ~, \nnb \\
B_1 &=& C_{10}\, g ~, \\
B_2 &=& C_{10} \, f ~, \nnb \\
F &=& 2 m_{\tau} C_{10} + \frac{m^2_B}{m^2_b} C_{Q_{2}} ~, \nnb \\
F_1 &=& \frac{m^2_B}{m^2_b} C_{Q_{1}}~.
\eea
In Eqs. (\ref{Msd}) and (\ref{Mib}),  $\varepsilon_\mu^\ast$ and $k_\mu$ 
are the four  vector polarization and four momentum of the photon, respectively, 
$q$ is the momentum transfer and $p_B$ is the momentum of the $B$ meson. 
The form factors $g$, $f$, $g_1$, $f_1$and $f_B$ are 
defined as follows \cite{Aliev1,Eilam2}:
\bea
\label{mel1}
\la \gamma(k) \vel \bar s \gamma_\mu
(1 \mp \gamma_5) b \ver B(p_B) \ra &=&
\frac{e}{m_B^2} \Big\{
\epsilon_{\mu\nu\lambda\sigma} \varepsilon^{\ast\nu} q^\lambda
k^\sigma g(q^2) \pm i\,
\Big[ \varepsilon^{\ast\mu} (k q) -
(\varepsilon^\ast q) k^\mu \Big] f(q^2) \Big\}~, \nnb 
\eea
\bea                                             
\la \gamma \ve \bar s i \sigma_{\mu \nu} k_\nu (1\mp \gamma_5) b \ve
B \ra &=& \frac{e}{m_B^2} \Bigg{\{} \epsilon_{\mu \alpha \beta \sigma}
\epsilon_\alpha^*
k_\beta q_\sigma \, g_1(p^2)
\mp ~ i \left[ \epsilon_\mu^* (k q) - (\epsilon^* k ) q_\mu \right] \,
f_1(p^2) \Bigg{\}}~,\label{ff2} \\
\la 0 \ve \bar s \gamma_\mu \gamma_5 b \ve B \ra &=& 
-~i f_B p_{B\mu}~. \nnb
\eea
In obtaining the expressions  (\ref{Msd}) and (\ref{Mib}), we have also used  
\bea
\la 0 \ve \bar s \sigma_{\mu\nu} (1+\gamma_5) b \ve B \ra &=& 0~,\nnb
\eea
and conservation of the vector current. Note that in contrast to 
${\cal M}_{SD}$ part of the amplitude, its ${\cal M}_{IB}$ part receives
contributions from NHB exchange diagrams, which are represented by  the 
factors $F$ and $F_1$ in Eq. (\ref{Mib}). During the calculations of these
NHB contributions in model III, we encountered logarithmic divergences and
used the on-shell renormalization scheme to overcome them. (For details, see
\cite{Ergur}).

We now examine the probability of the process  $B \rar \tau^+ \tau^- \gamma$  as a 
function of
the four momenta of the particles. In the center of mass (CM) frame of the
dileptons $\tau^+\tau^-$, where we take  $z=\cos \theta $ and $\theta$ is the angle
between the momentum of the $B$-meson and that of $\tau^-$, double differential decay width 
is found to be
\bea
\label{dGdxdz}
\frac{d \Gamma}{dx \, dz} = \frac{1}{(2 \pi)^3 64 }\, x \, v \, m_B \, \vel {\cal M} \ver^2~,
\eea 
with
\bea
\vel {\cal M} \ver^2 & = &\vel {\cal M}_{SD} \ver^2+\vel {\cal M}_{IB} \ver^2+
2 Re({\cal M}_{SD}{\cal M}^{*}_{IB} )\label{M2}
\eea
where $x=2 E_{\gamma}/m_B$ is the dimensionless photon energy and
$v=\sqrt{1-\frac{4 r}{1-x}}$ with $r=m^2_{\tau}/m^2_{B}$. After some
calculation, we get for the different parts of the squared matrix 
elements in Eq. (\ref{M2}) :
\bea
\label{Msd2}   
\vel {\cal M}_{SD} \ver^2 &=& \vel \frac{\alpha G_F }{2 \sqrt{2} \, \pi} 
V_{tb} V_{ts}^* \frac{e}{m^2_B} \ver^2 \, 
\Bigg( 8 \, {\rm Re} \ga A^*_2 B_1 + A^*_1 B_2 \dr p_B^2 \ga p_1 q - p_2 q \dr
\ga p_1 q + p_2 q \dr ~ \nnb \\
&& +~ 4 \left[ \vel B_1 \ver^2 + \vel B_2 \ver^2 \right] 
\left[ \ga p_B^2 - 2 m_\tau^2 \dr \ga \ga p_1 q \dr^2 + \ga p_2 q \dr^2
\dr - 4 m_\tau^2 \ga p_1 q \dr \ga p_2 q \dr \right]~  \nnb \\
&& +~ 4 \left[ \vel A_1 \ver^2 + \vel A_2 \ver^2 \right]
\Big[ \ga p_B^2 + 2 m_\tau^2 \dr \ga \ga p_1 q \dr^2 + \ga p_2 q \dr^2
\dr + ~ 4 m_\tau^2 \ga p_1 q \dr \ga p_2 q \dr \Big] \Bigg )~, \nnb \\ & & \\ 
2\, {\rm Re} \ga {\cal M}_{SD} {\cal M}_{IB}^* \dr &=&
~\vel \frac{\alpha G_F}{2 \sqrt{2} \, \pi} V_{tb} V_{ts}^* \frac{e}{m_B} \ver^2 \, f_B 
\Bigg{\{} 16 \, m_\tau \Bigg[ {\rm Re}(A_1 F^*) \, 
\frac{ \ga p_1 q +  p_2 q \dr^3}{\ga p_1 q \dr \ga p_2 q \dr}~ \nnb \\
&& +~ {\rm Re}(B_2 F^*) \,
\frac{ \ga p_1 q +  p_2 q \dr^2  \ga p_1 q -  p_2 q \dr}
{\ga p_1 q \dr \ga p_2 q \dr} \Bigg]   \nnb \\ 
&&-~\Bigg[ {\rm Re}(A_2 F^*_1) \, 
\frac{ \ga p_1 q +  p_2 q \dr^3}{\ga p_1 q \dr \ga p_2 q \dr} -{\rm Re}(B_1 F^*_1) \,
\frac{ \ga p_1 q +  p_2 q \dr^2  \ga p_1 q -  p_2 q \dr}
{\ga p_1 q \dr \ga p_2 q \dr} \Bigg] \nnb \\ 
&&+~\frac{m_{B}^{2}}{m_{b}} {\rm Re}(A_2 F^*_1) \Big[ \frac{(m_{\tau}^{2}-3 p_{2}q)
(p_{1}q)}{p_{2}q}+\frac{(2m_{\tau}^{2}- p_B^{2})(p_{2}q)}{p_{1}q}\Big] \Big{\}}~,
\eea
\bea
\label{Mib2}
\lefteqn{\vel {\cal M}_{IB} \ver^2 = -~\vel \frac{\alpha G_F}{2 \sqrt{2} \, \pi}
V_{tb} V_{ts}^* e f_B \ver^2 } \nnb \\
& & \Bigg{\{}4 ( \vel F \ver^{2}+\vel F_1 \ver^{2})\Bigg[ \frac{1}{p_1 q}
\left( 3 m_\tau^2 - p_B^2 -2 p_2 q \right)+\frac{1}{p_2 q}   
\left( 3 m_\tau^2 - p_B^2 -2 p_1 q \right)-4 \Bigg] \nnb \\ &+&
 \frac{2m_\tau^2}{\ga p_1 q \dr^2} \Bigg[ \vel F \ver^{2}\ga p_B^2 + 2 p_2 q \dr +
\vel F_1 \ver^{2}\ga p_B^2 + 2 p_2 q -4m_{\tau}^{2}\dr \Bigg] \nnb \\
&& +~ \frac{2m_\tau^2}{\ga p_2 q \dr^2} \Bigg[\ \vel F \ver^{2}\ga p_B^2 + 
2 p_1 q \dr + \vel F_1 \ver^{2}
\ga p_B^2 + 2 p_1 q -4m_{\tau}^{2}\dr \Bigg]~ \nnb \\
&& +\frac{2}{\ga p_1 q \dr \ga p_2 q \dr} \Bigg[ \vel F \ver^{2}p_B^2 \ga 
2m_{\tau}^{2}-p_B^2 \dr -
\vel F_1 \ver^{2}\ga p_B^2 + 2 p_2 q -4m_{\tau}^{2}\dr \Bigg] \Bigg{\}} ~. \nnb \\
& &
\eea

There is a singularity in $\vel {\cal M}_{IB} \ver^2$ at the lower limit of the photon 
energy due to
the soft photon emission from charged lepton line, while  $\vel {\cal M}_{SD}
\ver^2$  and $Re( {\cal M}_{SD} {\cal M}^*_{IB} )$ terms are free from this singularity. 
It has been shown that
when processes  $B \rar \tau^+ \tau^- \gamma$ and  $B \rar \tau^+ \tau^-$ are considered 
together,
the singular terms in $\vel {\cal M}_{IB} \ver^2$ exactly cancel the ${\cal O} (\alpha )$ 
virtual
correction in  $B \rar \tau^+ \tau^-$ amplitude. But instead of this approach we prefer 
the one 
used in ref.\cite{Aliev2} which amounts to impose a cut on the photon energy, i.e., we 
require 
$E_{\gamma}\geq 50 \,$ MeV. This restriction means that we only consider the
hard photons in the process $B \rar \tau^+ \tau^- \gamma $. Therefore, the
Dalitz boundary for the dimensionless photon energy is taken as 
\bea
\delta \leq x \leq 1-\frac{4 m_\tau^2}{ m^2_B}~, \label{KR}
\eea
with $\delta=0.01$.

Using Eqs. (\ref{dGdxdz})-(\ref{Mib2}), we get the following result for the
double differential decay width
\bea
\frac{d \Gamma}{dx dz} & = & \vel \frac{\alpha G_F}{2 \sqrt{2} \, \pi} V_{tb}
V_{ts}^*  \ver^2 \,
\frac{\alpha}{\ga 2 \, \pi \dr^3}\,\frac{\pi}{4}\,m_B \, x \, v \, \nnb \\
& & \Bigg\{ \frac{m^2_B}{32} \, x^2 \,
\Bigg[((1+z^2)(1-x-4 r)) ( \vel A_1 \ver^2 + \vel A_2 \ver^2 +\vel B_1 \ver^2 + \vel B_2 \ver^2 )
+8 r (\vel A_1 \ver^2 + \vel A_2 \ver^2 ) \nnb \\ 
& + & 4 z \sqrt{(1-x)(1-x-4 r)}\mbox{\rm Re}
(A_2 B^*_1+A_1 B^*_2)\Bigg ] \nnb \\ 
& + & f_B m_{\tau} \frac{(x-1)}{((z^2-1)(x-1)+4 r z^2)}\Bigg[ v x z 
\mbox{\rm Re} (B_2 F^*-B_1 F^*_1)+(1-4 r-z^2 (1-x-4 r))\nnb \\ 
& & \mbox{\rm Re}(A_2 F^*_1)-
x \mbox{\rm Re}(A_1 F^*)\Bigg ]+f^2_B \frac{(1-x)}{x^2 ((z^2-1)(x-1)+4 r z^2)^2}
\nnb \\
&& \Bigg [ \vel F \ver^2  \left((-2+4 x-3 x^2+x^3+8 r (1-x)) (z^2-1)+4 r
x^2 z^2  \right) \nnb \\ 
& + & \vel F_1 \ver^2 \left(\left(32 r^2 (x-1)+4 r (4-6 x+2 x^2)-2+4 x-3 x^2+x^3
\right)(z^2-1)+x^2 z^2 \right) \Bigg ] \Bigg\}\, . \nnb \\ & & 
\eea

Integrating the angle variable, we find the photon energy distribution given
by
\bea
\label{dGdx}
\frac{d\Gamma}{dx} & = &  \vel \frac{\alpha G_F}{2 \sqrt{2} \, \pi} 
V_{tb} V_{ts}^* \ver^2 \, \frac{\alpha}{\ga 2 \, \pi \dr^3}\,\frac{\pi}{4}\,m_B \, D(x)
\eea
where
\bea
D(x) & = & \frac{m_B^2 }{12} x^3 v\, \Bigg[ (\vel A_1 \ver^2 + \vel A_2 \ver^2
)(1+2 r-x)+
(\vel B_1 \ver^2 + \vel B_2 \ver^2 )(1-4 r-x)\Bigg] \nnb \\
& -& f_B m_{\tau} \, x \, \Bigg[ 2 v (1-x) \mbox{\rm Re}(A_2 F^*_1)+{\rm ln}
\frac{1 + v}{1-v}\Bigg( (x-4 r)\mbox{\rm Re} (A_2 F^*_1)-x \mbox{\rm Re}(A_1 F^*)  \Bigg )
\Bigg ]\nnb \\
& - & 2 f^2_B \Bigg[v \frac{(1-x)}{x} \Bigg( \vel F \ver^2+
(1-4 r) \vel F_1 \ver^2 \Bigg)
+ \, {\rm ln}\frac{1 + v}{1 - v} \Bigg(  \Bigg( 1 +\frac{2 r}{x} -
\frac{1}{x} -
\frac{x}{2} \Bigg)\vel F \ver^2 \nnb \\
& + & \Bigg(  (1-4 r) - \frac{2 \ga 1- 6 r + 8 r^2 \dr}{x}
 -\frac{x}{2}\Bigg) \vel F_1 \ver^2 \Bigg)\Bigg]  ~.\label{Dx}
\eea
We also give the forward-backward asymmetry, $A_{FB}$, in $B \rar \tau^+
\tau^-\gamma$. Using the definition of differential  $A_{FB}$
\begin{eqnarray} 
A_{FB}(x)& = & \frac{ \int^{1}_{0}dz \frac{d \Gamma }{dz} - 
\int^{0}_{-1}dz \frac{d \Gamma }{dz}}{\int^{1}_{0}dz 
\frac{d \Gamma }{dz}+ \int^{0}_{-1}dz \frac{d \Gamma }{dz}}
\label{AFB1} 
\end{eqnarray}
we find
\begin{eqnarray}
A_{FB}=\frac{\int\, dx\,E(x)}{\int\, dx\,D(x)}\, , 
\label{AFB2}
\end{eqnarray}
where
\begin{eqnarray}
E(x) & = & -4 \, v \, x^2 \Bigg( m^2_B \, x \, \sqrt{(x-1) (x-1+4 r)}\mbox{\rm Re}
(A_1 A^*_2-B_1 B^*_2) \nnb \\
& +& 4 f_B m_{\tau} \, v \, \Bigg(\frac{x-1}{x-1+4 r}\Bigg)
\ln \frac{4 r}{x-1}
\mbox{\rm Re}((A_2- B_2) F^*-(A_1- B_1) F^*_1) \Bigg ) \nnb \\ &&
\end{eqnarray}
and $D(x)$ is given by Eq.(\ref{Dx}).

Finally, we would like to discuss the $\tau^{-}$ lepton polarization
effects for the process  $B \rar \tau^+ \tau^- \gamma$. The longitudinal 
polarization asymmetry of the  $\tau^{-}$ lepton is defined as
\begin{eqnarray}
P_{L} (x) & = & \frac{(d\Gamma (S_L)/dx)-(d\Gamma (-S_L)/dx)}{(d\Gamma
(S_L)/dx)+(d\Gamma (-S_L)/dx)} \label{PL}
\end{eqnarray}
where $S_{L}$ is the orthogonal unit vector for the polarization of the
$\tau^{-}$ lepton to the longitudinal direction (L) and in the CM frame of
the $\tau^+\tau^-$ system, it is defined as
\begin{eqnarray}
S^{\mu}_{L} & = & \Bigg(\frac{|\vec{p}_1|}{m_{\tau}}\, , \,
\frac{E_{\tau}\vec{p}_1}{m_{\tau}|\vec{p}_1|}\, \Bigg).
\end{eqnarray} 
Here, $\vec{p}_1$ and $E_{\tau}$ are  the three momentum and energy
of the $\tau^{-}$ lepton in the CM frame, respectively. Calculation of 
$P_L$ leads to the following result
\begin{eqnarray}
P_L & = & \frac{2}{3}\frac{1}{v D(x)}\Bigg\{ -m^2_B v^3 x^2 (-1+x)^2 
(\vel A_1 \ver^2 + \vel A_2 \ver^2-\vel B_1 \ver^2 - \vel B_2 \ver^2 ) 
\nnb \\ & + & 12 f^2_B \frac{1}{(-1+x)^2}(1+x^2-4 r (1+x)) 
\Bigg(x v +(2 r-x) {\rm ln}\frac{1 + v}{1 - v}\Bigg) \, 
\mbox{\rm Re}(F F^*_1)\nnb \\
& + & 6 f_B m_{\tau} \Bigg[ \Bigg(v x-2 r {\rm ln}\frac{1 + v}{1 - v} \Bigg)
((1+x) \,  \mbox{\rm Re}((A_2+B_2) F^*)+(-1+x) \,\mbox{\rm Re}((A_1+B_1)
F^*_1))\nnb \\ & + & (-1+x) 
\Bigg(x v +(2 r-x) {\rm ln}\frac{1 + v}{1 -v}\Bigg)\,
\mbox{\rm Re}((A_2-B_2) F^*_1-(A_1-B_1)F^*)  \Bigg] \Bigg\}
\end{eqnarray}

In order to investigate the dependence of the $\tau^-$ lepton  polarization 
on the model III parameters, we eliminate the other parameter, namely $x$,
by performing the $x$-integrations over the allowed kinematical region
(Eq.(\ref{KR})) so
as to obtain the  averaged lepton polarization. For the longitudinal component  
the averaged lepton polarization  is defined as
\begin{eqnarray}
<P_L> & = & \frac{\int_{\delta}^{(1-4 m^2_{\tau}/m^2_B)} P_L \,
\frac{d\Gamma}{dx} \, dx}{\int_{\delta}^{(1-4 m^2_{\tau}/m^2_B)}
\frac{d\Gamma}{dx} \, dx} \nnb .
\end{eqnarray}

For the process  $B \rar \tau^+ \tau^- \gamma$, the lepton polarization has,
in addition to the longitudinal component $P_L$, transverse and normal
components. Since these two orthogonal components are proportional to the tau lepton 
mass, they are expected to be  significant for the $\tau^+\tau^-$ channel.
We shall discuss their effects  in a more detailed paper.  

\section{Numerical analysis and discussion}
To calculate the decay width, first of all, we need the 
explicit forms of the form factors $g,~f,~g_1$ and $f_1$. In refs.
\cite{Buchalla} and \cite{Eilam2}, 
they are calculated in the framework of light--cone QCD sum rules 
and their $q^2$ dependences, to a very good accuracy,
can be represented in the following dipole
forms,
\bea
\label{ff}
g(q^2) &=& \frac{g(0)}{\left(1-\frac{q^2}{m_g^2}\right)^2~},
~~~~~~~~~~
f(q^2) = \frac{f(0) }{\left(1-\frac{q^2}{m_f^2}\right)^2}~,
 \nnb \\
g_1(q^2) &=& \frac{g_1(0)}{\left(1-\frac{q^2}{m^2_{g_1}}\right)^2}~,
~~~~~~~~~
f_1(q^2) = \frac{f_1(0)}{\left(1-\frac{q^2}{m^2_{f_1}}\right)^2}~,
\eea
where
\bea
g(0) & = & 1 \, GeV \, , \,f(0)=0.8 \, GeV \, , \, g_1(0)=3.74 \, GeV^2 \, , 
\,f_1(0)=0.68 \, GeV^2 \,, \nnb \\ 
m_g & = & 5.6 \, GeV \, , \, m_f=6.5 \, GeV \, , \, m_{g_1}=6.4 \, GeV \, , \, 
m_{f_1}=5.5 \, GeV \, . \nnb 
\eea
In addition to these form factors, the other input parameters which we have used 
in our numerical calculations are given in table I.

For the free parameters of the model III, namely,  the masses 
of charged and neutral Higgs bosons, $m_{H^{\pm}}, m_{A^0}, m_{h^0},
m_{H^0}$ and the Yukawa couplings ($\xi_{ij}^{U,D}$),
we use  the restrictions coming from $B\rightarrow X_s \, \gamma $ decay,
whose BR is given by CLEO measurement \cite{CLEO} as 
\bea
\label{cleomes}
BR(B \rar X_s \, \gamma ) & = & (3.15 \pm 0.35 \pm 0.32 ) \times 10^{-4}
\eea
and $B^0-\bar{B}^0$ mixing \cite{Aliev4}, $\rho $ parameter \cite{Atwood} and 
neutron electric-dipole moment \cite{D.Bowser}, that yields  
$\bar{\xi}^{D}_{N, ib} \sim 0$ and $\bar{\xi}^{D}_{N, ij}\sim 0$, 
where the indices $i,j$ denote d and s quarks, and $\bar{\xi}_{N tc} <<
\bar{\xi}^{U}_{N tt}$. Therefore,   we take 
into account only the Yukawa couplings of b and t  quarks, $\bar{\xi}^{U}_{N,tt}$ and
$\bar{\xi}^{D}_{N,bb}$ and also $\bar{\xi}^{D}_{N,\tau \tau}$. Further,
in our numerical calculations we  adopted the restriction,  $0.257 \leq
|C_7^{eff}| \leq 0.439$
due to the CLEO measurement, Eq.(\ref{cleomes}), (see \cite{Aliev4} for details)
and  the redefinition
\begin{eqnarray}
\xi^{U,D}=\sqrt{\frac{4 G_F}{\sqrt{2}}} \bar{\xi}^{U,D} \nonumber \,\, .
\label{xineutr}
\end{eqnarray}

Before we present our results, a small note about the calculations 
of the long distance (LD) effects is in
place.  We take into account five possible
resonances for the LD effects coming from the reaction $b \rar s \, \psi_i
\rar s \, \tau^+ \tau^-$, where $i=1,...,5$ and divide the integration
region into two parts: $\delta \leq x \leq 1-((m_{\psi_2}+0.02)/m_B)^2$ and
$1-((m_{\psi_2}-0.02)/m_B)^2 \leq x \leq 1-(2 m_{\tau}/m_B)^2 $, where 
$m_{\psi_2}=3.686 $ GeV is the mass of the second resonance. (See Appendix
for the details of LD contributions). 

\begin{figure}[htb]
\vskip -3.0truein
\centering
\epsfxsize=6.8in
\leavevmode\epsffile{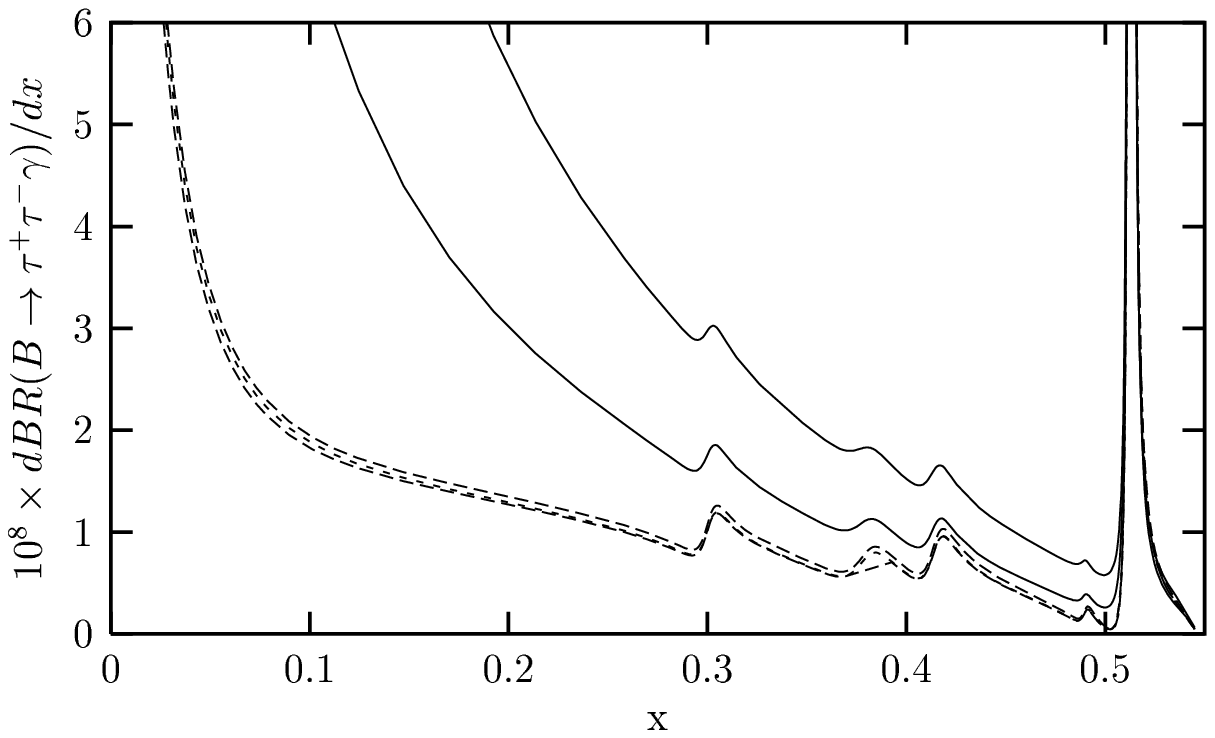}
\vskip -3.0truein
\caption[]{Differential BR  as a function of $x$ for $\bar{\xi}_{N,bb}^{D}=40\, m_b$
and $\bar{\xi}_{N,\tau\tau}^{D}=10\, m_{\tau}$, in case of the ratio
$ |r_{tb}| <1$. }
\label{dGNHBLDa}
\end{figure}

\begin{figure}[htb]
\vskip -3.0truein
\centering
\epsfxsize=6.8in
\leavevmode\epsffile{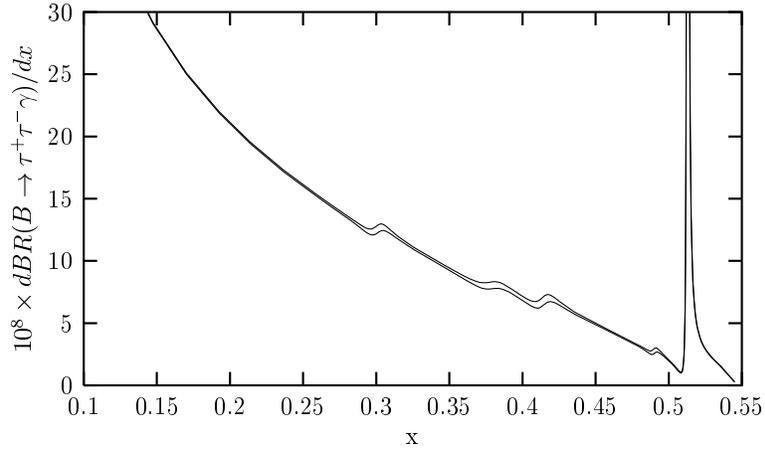}
\vskip -3.0truein
\caption[]{The same as Fig.(\ref{dGNHBLDa}), but for $ r_{tb} >1$ with 
$\bar{\xi}_{N,bb}^{D}=0.1\, m_b$ and $\bar{\xi}_{N,\tau\tau}^{D}=
m_{\tau}$.}
\label{dGNHBLDb}
\end{figure}

\begin{figure}[htb]
\vskip -3.0truein
\centering
\epsfxsize=6.8in
\leavevmode\epsffile{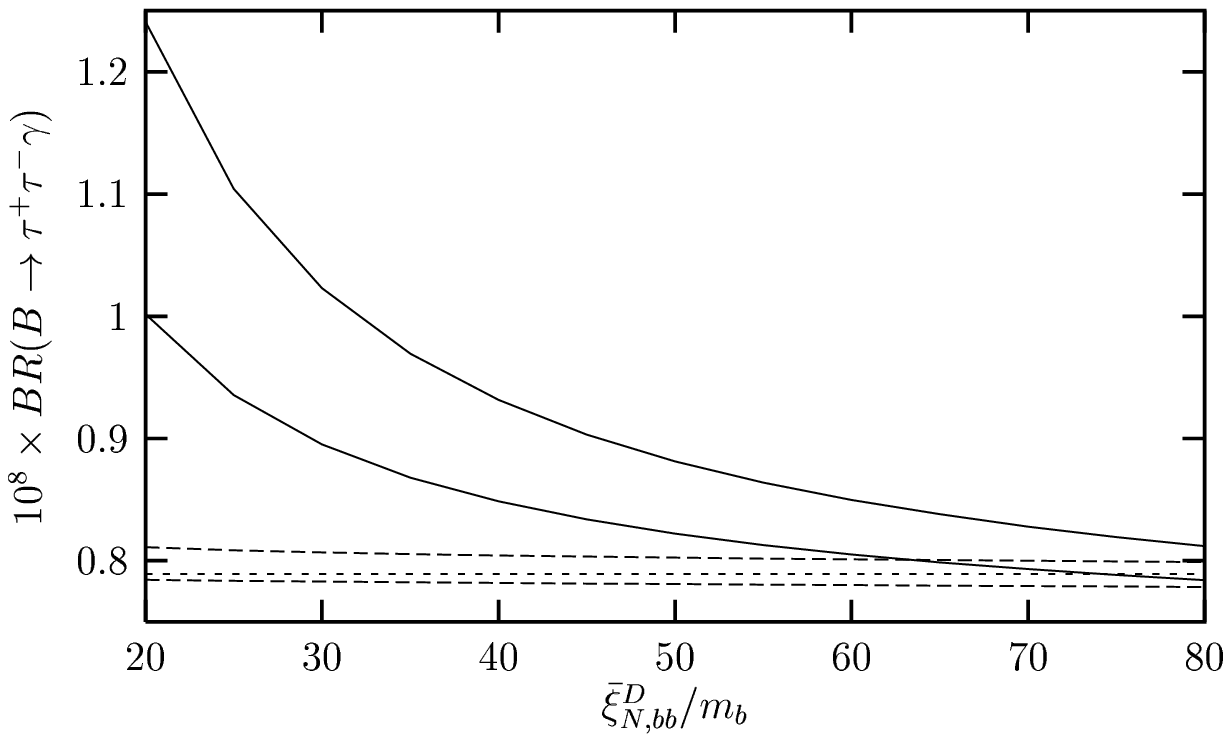}
\vskip -3.0truein
\caption[]{BR as a function of  $\bar{\xi}_{N,bb}^{D}/m_{b}$  for
$\bar{\xi}_{N,\tau \tau}^{D}= m_{\tau}$, 
in case of the ratio $ |r_{tb}| <1$. }
\label{GamIIIkbbrk1b}
\end{figure}

\begin{figure}[htb]
\vskip -3.0truein
\centering
\epsfxsize=6.8in
\leavevmode\epsffile{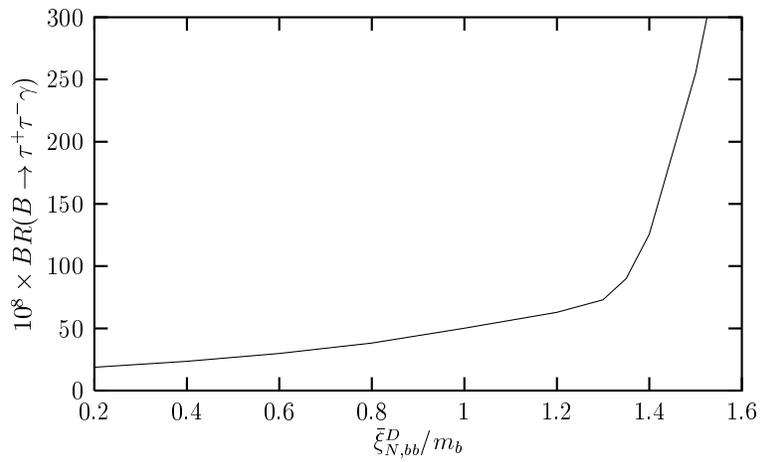}
\vskip -3.0truein
\caption[]{The same as Fig.(\ref{GamIIIkbbrk1b}), but for $ r_{tb} >1$.}
\label{GamIIIkbbrb1n}
\end{figure}

\begin{figure}[htb]
\vskip -3.0truein
\centering
\epsfxsize=6.8in
\leavevmode\epsffile{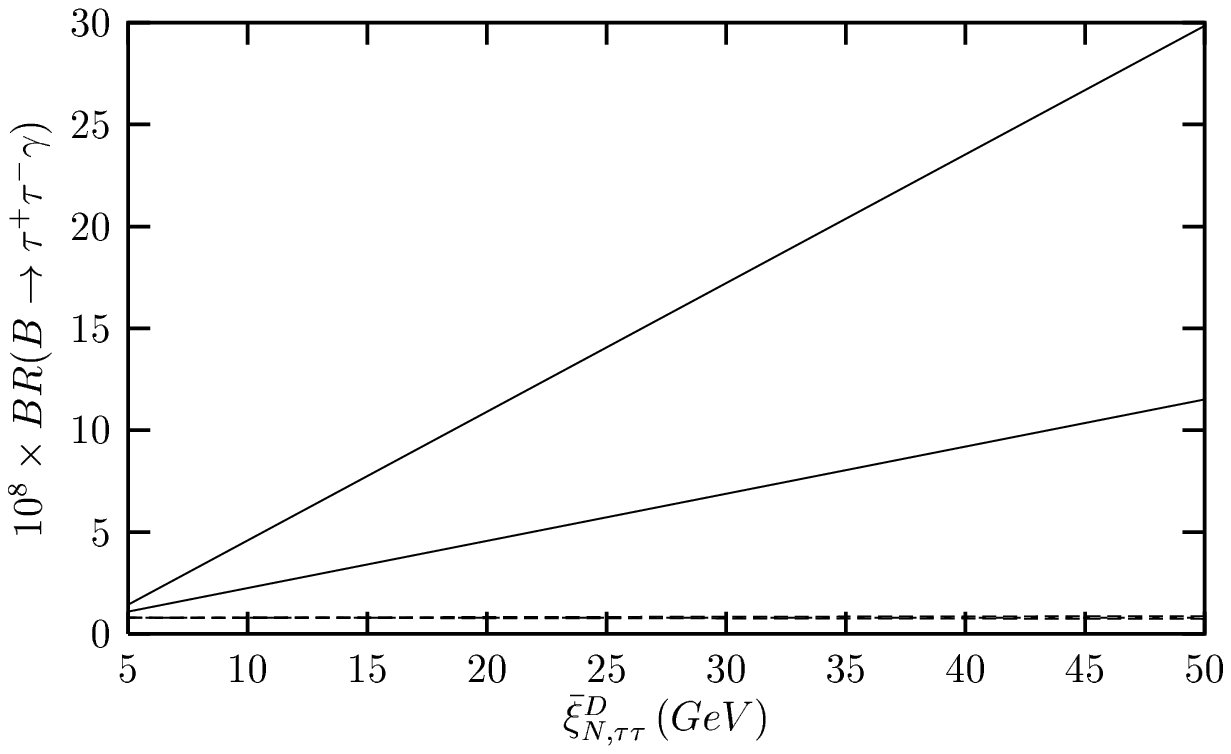}
\vskip -3.0truein
\caption[]{BR as a function of   $\bar{\xi}_{N,\tau\tau}^{D}$
for  $\bar{\xi}_{N,bb}^{D}=40\,m_b$ and $|r_{tb}| <1$. }
\label{GamIIIkttrk1}
\end{figure}

\begin{figure}[htb]
\vskip -3.0truein
\centering
\epsfxsize=6.8in
\leavevmode\epsffile{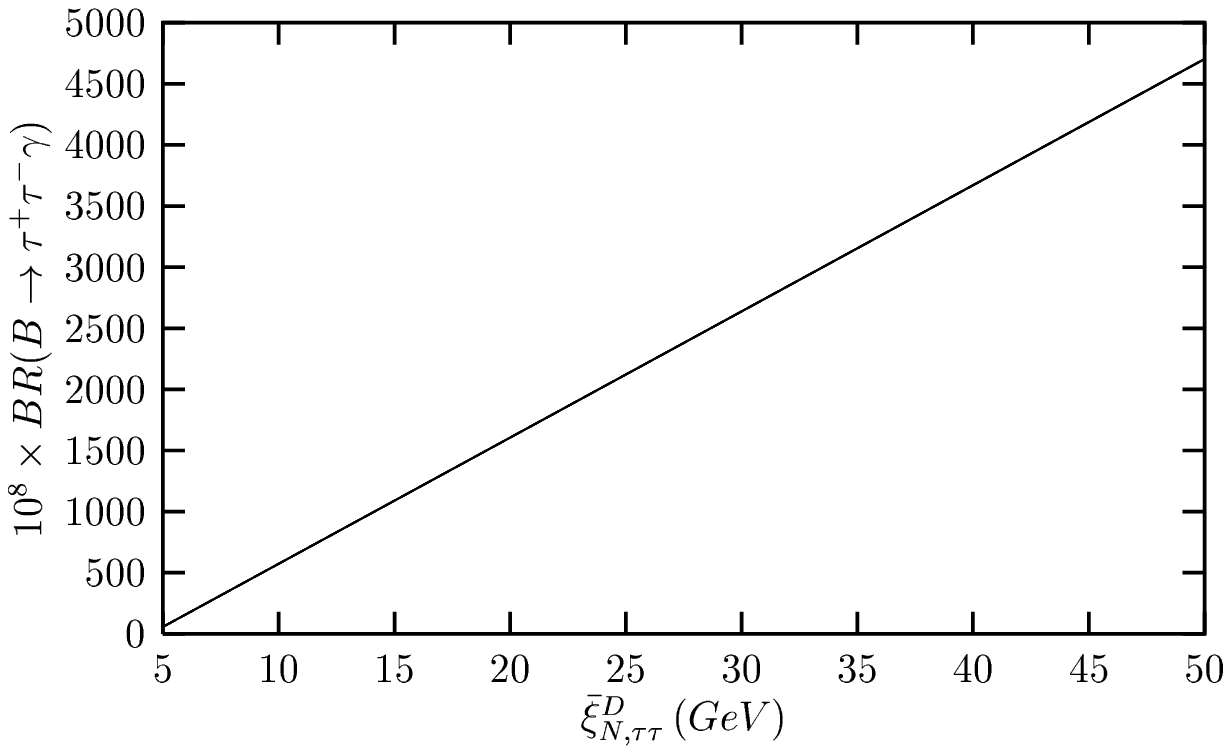}
\vskip -3.0truein
\caption[]{The same as Fig. (\ref{GamIIIkttrk1}), but for
$\bar{\xi}_{N,bb}^{D}=0.1\,m_b$ and $r_{tb} >1$.}
\label{GamIIIkttrb1}
\end{figure}

\begin{figure}[htb]
\vskip -3.0truein
\centering
\epsfxsize=6.8in
\leavevmode\epsffile{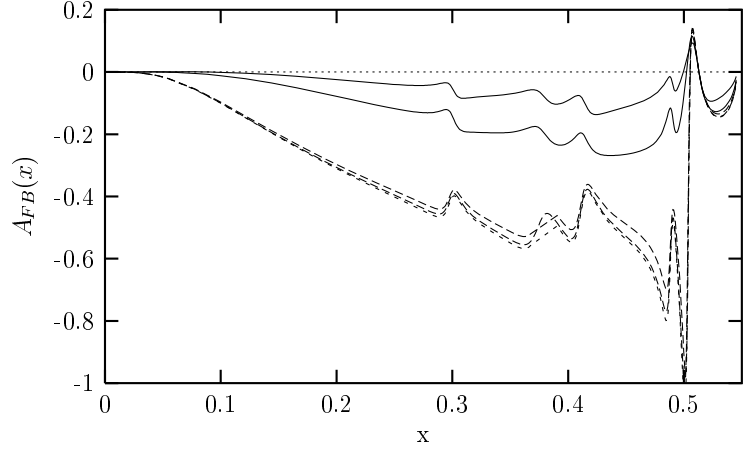}
\vskip -3.0truein
\caption[]{Differential $A_{FB}(x)$ as a function of $x$ for
$\bar{\xi}_{N,\tau\tau}^{D}=10 \,m_{\tau}$,
$\bar{\xi}_{N,bb}^{D}=40\,m_b$ and $|r_{tb}| <1$. }
\label{dAFBdxrk1}
\end{figure}

\begin{figure}[htb]
\vskip -3.0truein
\centering
\epsfxsize=6.8in
\leavevmode\epsffile{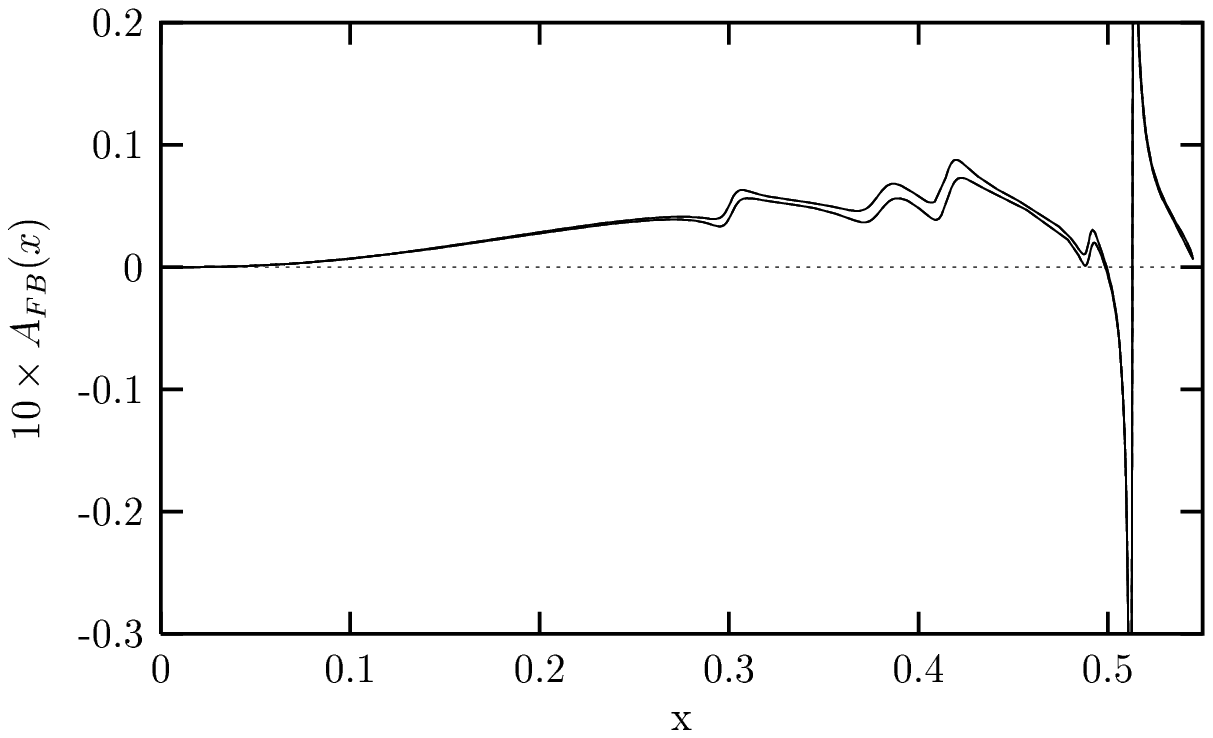}
\vskip -3.0truein
\caption[]{The same as Fig.( \ref{dAFBdxrk1}) but for
$\bar{\xi}_{N,bb}^{D}=0.1\,m_b$ and $r_{tb} >1$.}
\label{dAFBdxrb1}
\end{figure}

\begin{figure}[htb] 
\vskip -3.0truein   
\centering
\epsfxsize=6.8in    
\leavevmode\epsffile{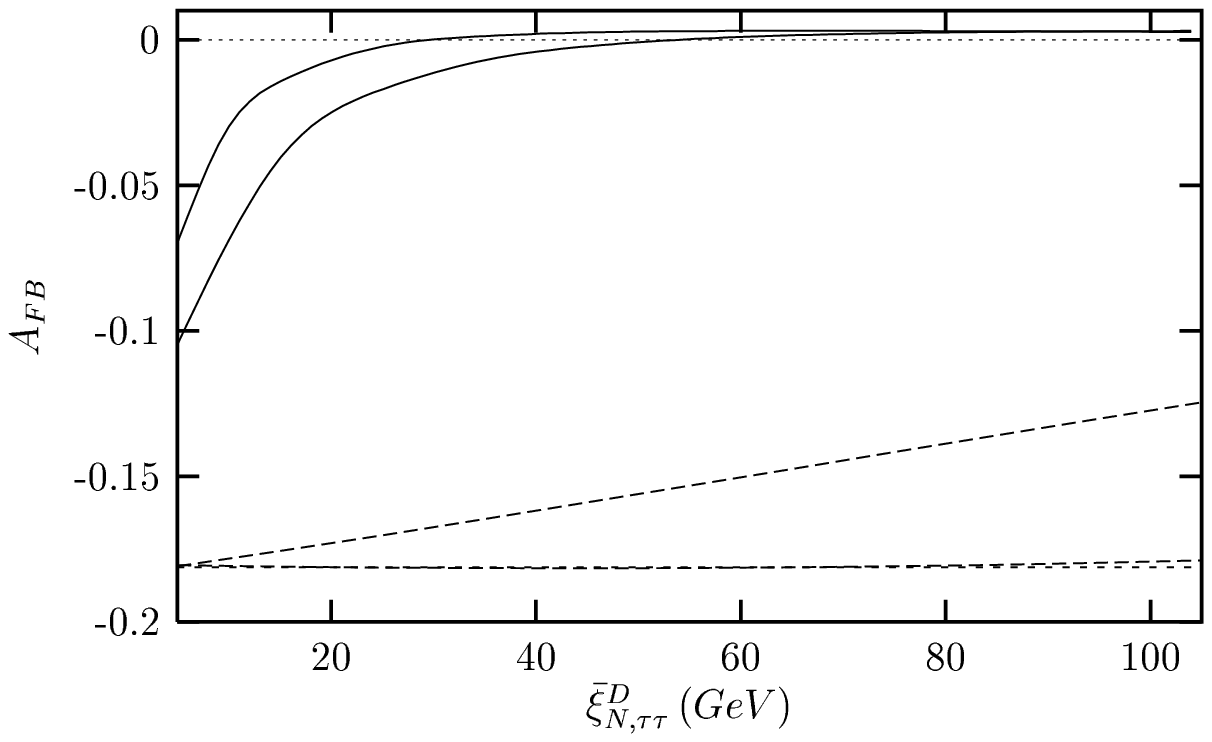}
\vskip -3.0truein   
\caption[]{$A_{FB}$ as a function of $\bar{\xi}_{N,\tau\tau}^{D}$ for
$\bar{\xi}_{N,bb}^{D}=40\,m_b$ and $|r_{tb}| <1$. }
\label{AFBIIIkttrk1}    
\end{figure}

\begin{figure}[htb]
\vskip -3.0truein
\centering
\epsfxsize=6.8in
\leavevmode\epsffile{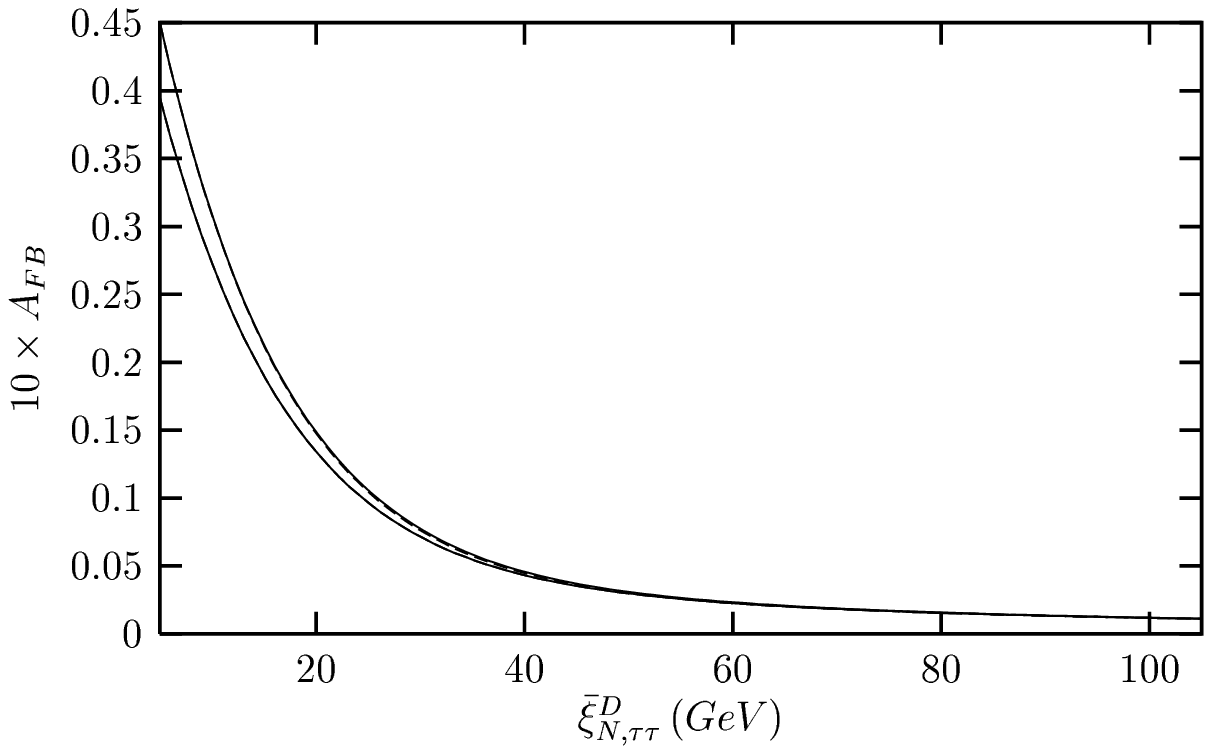}
\vskip -3.0truein
\caption[]{The same as Fig. (\ref{AFBIIIkttrk1}), but for
$\bar{\xi}_{N,bb}^{D}=0.1\,m_b$ and $r_{tb}>1$.}
\label{AFBIIIkttrb1a} 
\end{figure}

In this section, we first study the dimensionless photon energy dependence of the 
differential branching ratio $(d BR/dx)$ and the model III parameters
dependence of the BR and also the forward-backward asymmetry, $A_{FB}$.
The results of our calculations are presented through the graphs in
Fig.(\ref{dGNHBLDa})-(\ref{AFBIIIkttrb1a}).
In Fig.(\ref{dGNHBLDa}), we present $dBR (B\rightarrow \tau^+\tau^- \gamma )/dx $ as a
function of $x=2 E_{\gamma}/m_{B}$  for
$\bar{\xi}_{N,bb}^{D}=40\, m_b$ and 
$\bar{\xi}_{N,\tau\tau}^{D}=10\, m_{\tau}$, in case of the ratio
$ |r_{tb}|\equiv |\frac{\bar{\xi}_{N,tt}^{U}}{\bar{\xi}_{N,bb}^{D}}| <1$,
including  the long distance contributions. Here, the differential BR lies
in the region bounded by dashed (solid) curves for $C^{eff}_7 <0$
($C^{eff}_7 >0$). We see from this figure that there is an enhancement for
the $dBR (B\rightarrow \tau^+\tau^- \gamma )/dx $ in model III compared to
the SM result for the  $C^{eff}_7 >0$ case, while for  $C^{eff}_7 <0$, model
III predictions almost coincide with the SM one (small dashed curve). 
Fig. (\ref{dGNHBLDb}) is devoted the same
dependence of differential BR, but for $\bar{\xi}_{N,bb}^{D}=0.1\, m_b$ and 
$\bar{\xi}_{N,\tau\tau}^{D}= m_{\tau}$, in case of the ratio
$ r_{tb} >1$. We see that model III predictions for the  $C^{eff}_7 >0$ and
$C^{eff}_7 <0$  almost coincide and  they are  
one order larger  compared to both $ |r_{tb}| <1$  case and the SM one.

Fig (\ref{GamIIIkbbrk1b}) and (\ref{GamIIIkbbrb1n}) show  
$\bar{\xi}_{N,bb}^{D}/m_{b}$ dependence of BR  for
$\bar{\xi}_{N,\tau \tau}^{D}= m_{\tau}$ ,
in case of the ratio $ |r_{tb}| <1$ and  $ r_{tb} >1$, respectively. In
Fig.(\ref{GamIIIkbbrk1b}), BR is restricted in the region between  dashed lines 
(solid curves)  for
$C^{eff}_7 <0$ ($C^{eff}_7 >0$), while the small dashed straight line shows
the SM contribution. In Fig. (\ref{GamIIIkbbrb1n}), there is a  single curve since 
the    contributions  for both
$C^{eff}_7 >0$ and $C^{eff}_7 <0$ fit onto each other. We see that BR is
quite sensitive  to the parameter $\bar{\xi}_{N,bb}^{D}/m_{b}$, for both  
$|r_{tb}| <1$ and  $ r_{tb} >1$, however the  behavior
is opposite for  these two cases; for $|r_{tb}| <1$, BR is decreasing
with the increasing values of $\bar{\xi}_{N,bb}^{D}/m_{b}$, while for 
$r_{tb} >1$, it is increasing. Further, BR is 2-3 orders larger compared
to the SM result for $r_{tb} >1$ case.  For $|r_{tb}| <1$,  the enhancement 
with respect to the SM prediction is relatively moderate; nearly $(30-40)\%$
for $C^{eff}_7 >0$, but for $C^{eff}_7 <0$ they almost coincide with the SM one.

The dependence of the BR on the Yukawa coupling   $\bar{\xi}_{N,\tau\tau}^{D}$   
is presented in Fig. (\ref{GamIIIkttrk1}) ((\ref{GamIIIkttrb1})), 
for $|r_{tb}| <1$ (  $ r_{tb} >1$) case with
$\bar{\xi}_{N,bb}^{D}=40\, m_b$ ($\bar{\xi}_{N,bb}^{D}=0.1\, m_b$). 
It is seen that BR is increasing with the increasing values of 
$\bar{\xi}_{N,\tau\tau}^{D}$ and this is the contribution due to NHB
effects.  
From Fig. (\ref{GamIIIkttrk1}), we see that the BR lies in the region bounded 
by solid lines 
for $C^{eff}_7 >0$ and it is sensitive to the NHB effects, while  for
$C^{eff}_7 <0$, it is almost the same as the SM result (dashed straight
line). Note that the SM prediction for the  BR
is $7.89\times 10^{-9}$ and in model III without NHB effects, when $ |r_{tb}|
<1$ it is in  between $(7.83-7.97)\times 10^{-9}$  for  $C^{eff}_7 >0$ and  
$(8.73-8.50)\times 10^{-9}$ for $C^{eff}_7 <0$. When  $ r_{tb} >1$,
upper and lower limits of the BR without NHB effects are  
$(1.03-1.04)\times 10^{-7}$ for both  $C^{eff}_7
>0$ and  $C^{eff}_7 <0$. Thus, contribution from NHB effects is
seen  to reach the values that are   two orders of magnitude larger than the overall
contributions  for both $|r_{tb}| <1$ and  $ r_{tb}>1$, even for  small
values of $\bar{\xi}_{N,\tau\tau}^{D}$.

In Fig. (\ref{dAFBdxrk1}), the differential $A_{FB}(x)$ is shown for
$\bar{\xi}_{N,bb}^{D}=40\, m_b$ and
$\bar{\xi}_{N,\tau\tau}^{D}=10\, m_{\tau}$, in case of the ratio
$ |r_{tb}| <1$. Here, $A_{FB}(x)$ is restricted in the region between solid 
curves for $C^{eff}_7 >0$. 
It is seen that the value of  $|A_{FB}(x)|$ stands less than the SM one.
The dashed curves represent  
$C^{eff}_7 <0$ case and they almost coincide with the SM prediction for $A_{FB}(x)$.
Fig.(\ref{dAFBdxrb1})
is the same as Fig. (\ref{dAFBdxrk1}), but for $ r_{tb} >1$ with
$\bar{\xi}_{N,bb}^{D}=0.1\, m_b$ . For this case, the sign of $A_{FB}(x)$ is
opposite to the SM prediction and  $|A_{FB}(x)|$ is one order of magnitude smaller 
than the SM one.

In Fig. (\ref{AFBIIIkttrk1}) we plot the $A_{FB}$ as a function of the Yukawa coupling
$\bar{\xi}_{N,\tau\tau}^{D}$ for $\bar{\xi}_{N,bb}^{D}=40\, m_b$ and
$ |r_{tb}| <1$. $A_{FB}$  lies in the region bounded by dashed (solid) lines
for $C^{eff}_7 <0$ ( $C^{eff}_7 >0$ ). As seen from 
Fig. (\ref{AFBIIIkttrk1}) that $A_{FB}$ vanishes for the large values of 
$\bar{\xi}_{N,\tau\tau}^{D}$  for $C^{eff}_7 >0$, while $C^{eff}_7 <0$,
it does not vanish in the given region of $\bar{\xi}_{N,\tau\tau}^{D}$.
Contributions to $|A_{FB}|$ from model III stand less than the SM ones.
Fig.(\ref{AFBIIIkttrb1a}) is the same as  Fig. (\ref{AFBIIIkttrk1}) , but
for   $ r_{tb} >1$ with $\bar{\xi}_{N,bb}^{D}=0.1\, m_b$. Here,
contributions for $C^{eff}_7 >0$  and $C^{eff}_7 <0$  coincide and both
are restricted by the solid curves. 

\begin{figure}[htb]
\vskip -3.0truein
\centering
\epsfxsize=6.8in  
\leavevmode\epsffile{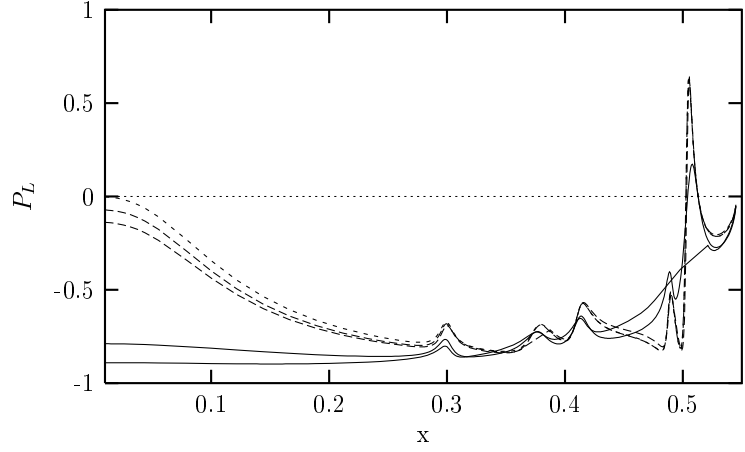}
\vskip -3.0truein
\caption[]{$P_L$  as a function of $x$ for
$\bar{\xi}_{N,bb}^{D}=40\, m_b$
and $\bar{\xi}_{N,\tau\tau}^{D}=10\, m_{\tau}$, in case of the ratio
$ |r_{tb}| <1$.}
\label{PLxIIINHBrk1}
\end{figure}

\begin{figure}[htb]
\vskip -3.0truein
\centering
\epsfxsize=6.8in
\leavevmode\epsffile{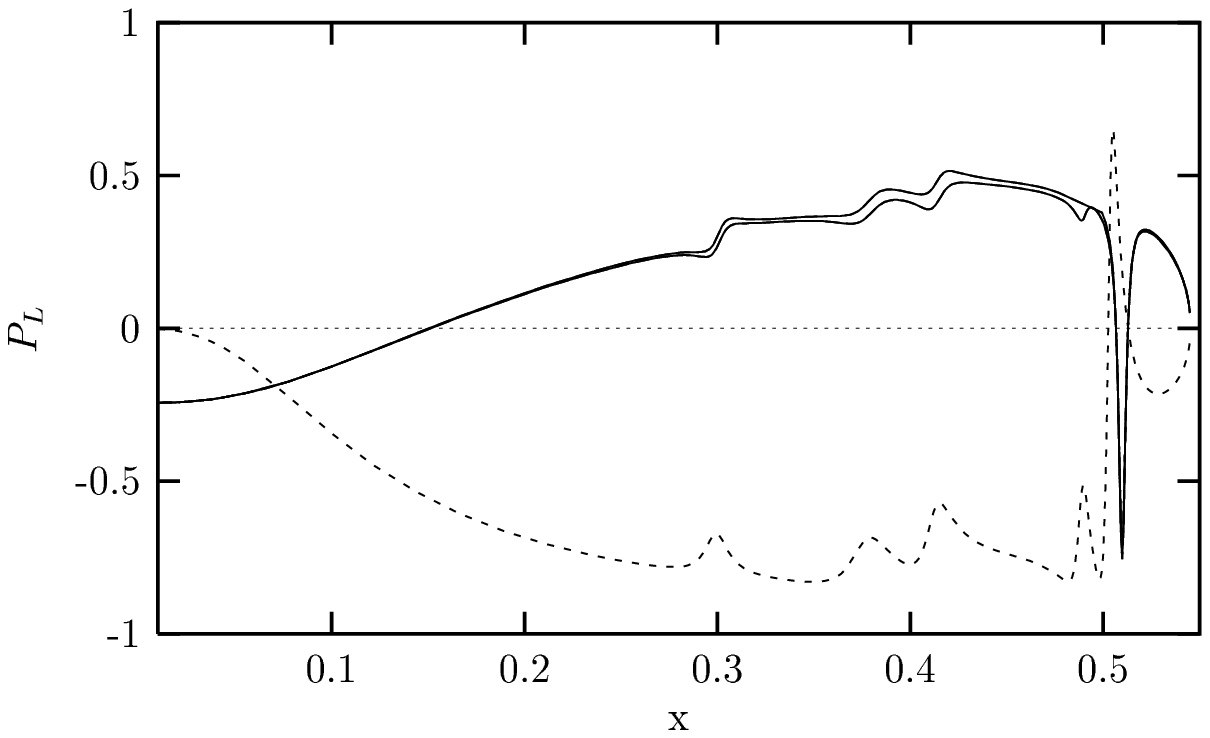}
\vskip -3.0truein
\caption[]{The same as Fig. (\ref{PLxIIINHBrk1}), but for
$\bar{\xi}_{N,bb}^{D}=0.1\,m_b$ and $r_{tb}>1$.}
\label{PLxIIINHBrb1}
\end{figure}

\begin{figure}[htb]
\vskip -3.0truein
\centering
\epsfxsize=6.8in  
\leavevmode\epsffile{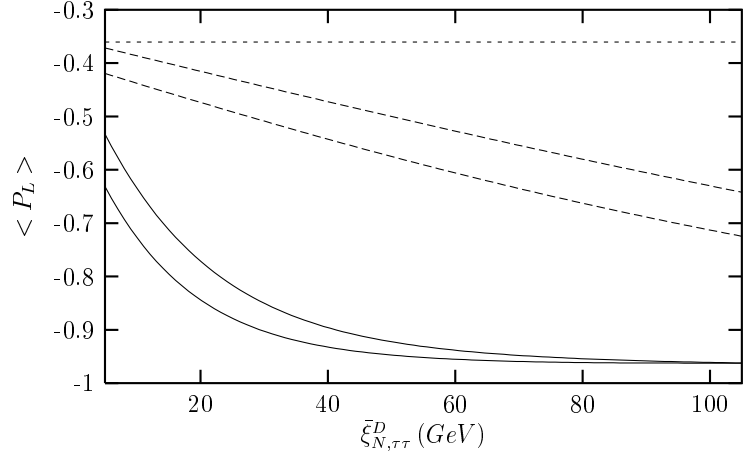}
\vskip -3.0truein
\caption[]{$<P_L>$  as a function of $\bar{\xi}_{N,\tau\tau}^{D}$ for
$\bar{\xi}_{N,bb}^{D}=40\, m_b$, in case of the ratio
$ |r_{tb}| <1$.}
\label{PLIIIkttrk1}
\end{figure}

\begin{figure}[htb]
\vskip -3.0truein
\centering       
\epsfxsize=6.8in
\leavevmode\epsffile{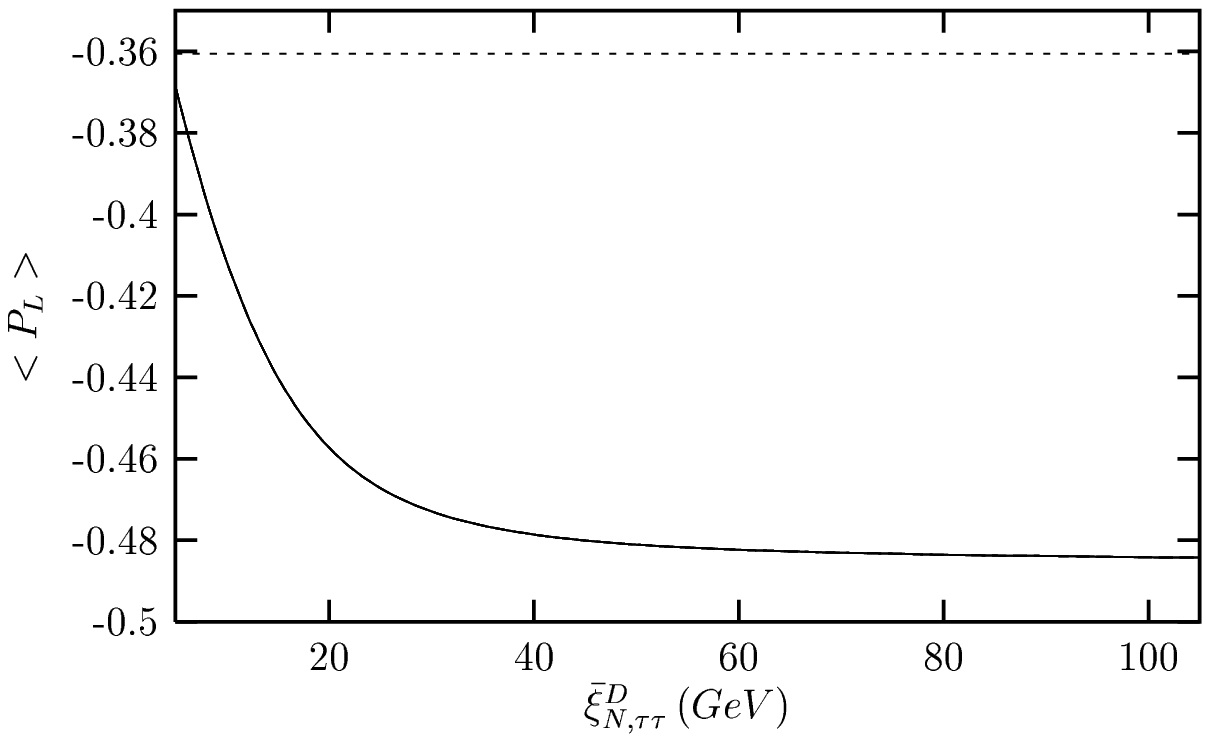}
\vskip -3.0truein
\caption[]{The same as Fig. (\ref{PLIIIkttrk1}), but for
$\bar{\xi}_{N,bb}^{D}=0.1\,m_b$ and $r_{tb}>1$.}
\label{PLIIIkttrb1}
\end{figure}

We present our analysis on the longitudinal component of the $\tau^-$ lepton
polarization through the graphs in
Figs.(\ref{PLxIIINHBrk1})-(\ref{PLIIIkttrb1}).
The dependence of $P_L$ on $x$ is presented in 
Fig (\ref{PLxIIINHBrk1}), for $ |r_{tb}| <1$ case with   
$\bar{\xi}_{N,\tau \tau}^{D}= 10 \, m_{\tau}$.  Here, 
$P_L$ is restricted in the region between  dashed  (solid ) curves  for
$C^{eff}_7 <0$ ($C^{eff}_7 >0$), while the small dashed curve shows
the SM contribution. From this figure, we see that the 2HDM contributions
change  $P_L$ significantly compared to the SM case for $C^{eff}_7 >0$, 
especially  for the small values of $x$. Fig. (\ref{PLxIIINHBrb1}) is the same as
Fig. (\ref{PLxIIINHBrk1}), but  $ r_{tb} >1$ with $\bar{\xi}_{N,bb}^{D}=0.1\,
m_b$. In this figure, two solid curves restrict the possible values of
$P_L$ for both $C^{eff}_7 <0$ and $C^{eff}_7 >0$ and it is seen that both 
the magnitude and the sign of $P_L$ are changed for  $ r_{tb} >1$.

The dependence of  $<P_L>$  on the NHB parameter 
$\bar{\xi}_{N,\tau\tau}^{D}$  is presented in Fig. (\ref{PLIIIkttrk1}) 
((\ref{PLIIIkttrb1})), for $|r_{tb}| <1$ (  $ r_{tb} >1$) case with
$\bar{\xi}_{N,bb}^{D}=40\, m_b$ ($\bar{\xi}_{N,bb}^{D}=0.1\, m_b$).
We see that for
$|r_{tb}| <1$,   $<P_L>$ lies in the region bounded by dashed lines 
(solid curves) for $C^{eff}_7 <0$ ( $C^{eff}_7 >0$) (Fig.
(\ref{PLIIIkttrk1})), while in case of  $ r_{tb} >1$, the contributions for
both $C^{eff}_7 <0$  and $C^{eff}_7 >0$ coincide and they are represented by the
single solid curve in Fig.(\ref{PLIIIkttrb1}). It is obvious that $<P_L>$ is 
very much sensitive to the NHB effects for both $|r_{tb}| <1$ and  $ r_{tb} >1$ 
cases. We note that SM prediction for $<P_L>$ is $-0.36$ and in model III
without NHB effects, when 
$|r_{tb}| <1$, it is about $-0.31$ for  $C^{eff}_7 >0$ and
$-0.36$  for $C^{eff}_7 <0$. When  $ r_{tb} >1$, the value of $<P_L>$ without 
NHB effects is about $-0.36$ for both  $C^{eff}_7>0$ and  $C^{eff}_7 <0$. 
Thus,  the value of $<P_L>$ without NHB effects reaches at
most the SM prediction, but NHB effects enhance it between $(5-100)\%$, even 
for  small values of $\bar{\xi}_{N,\tau\tau}^{D}$.
\begin{table}[h]
        \begin{center}
        \begin{tabular}{|l|l|}
        \hline
        \multicolumn{1}{|c|}{Parameter} & 
                \multicolumn{1}{|c|}{Value}     \\
        \hline \hline
        $m_{\tau}$                   & $1.78$ (GeV) \\
        $m_c$                   & $1.4$ (GeV) \\
        $m_b$                   & $4.8$ (GeV) \\
        $\alpha_{em}^{-1}$      & 129           \\
        $\lambda_t$            & 0.04 \\
        $m_{t}$             & $175$ (GeV) \\
        $m_{W}$             & $80.26$ (GeV) \\
        $m_{Z}$             & $91.19$ (GeV) \\
        $m_{H^0}$             & $150$ (GeV) \\
        $m_{h^0}$             & $70$ (GeV) \\  
        $m_{A^0}$             & $80$ (GeV) \\  
        $m_{H^{\pm}}$         & $400$ (GeV) \\
        $\Lambda_{QCD}$             & $0.225$ (GeV) \\
        $\alpha_{s}(m_Z)$             & $0.117$  \\
        $sin^2 \theta_W$             & $0.2325$  \\
        \hline
        \end{tabular}
        \end{center}
\caption{The values of the input parameters used in the numerical
          calculations.}
\label{input}
\end{table}
We would like to summarize our results:
\begin{itemize}
\item  We observe an enhancement in the differential branching ratio
and branching ratio for the exclusive process $B \rar \tau^+ \tau^- \gamma$
in the general 2HDM compared to the SM predictions. For $|r_{tb}|<1$ case, this
enhancement  is much more detectable for $C_{7}^{eff}>0$
case compared to the $C_{7}^{eff}<0$ one. For $r_{tb} >1$, we see that
contributions for $C_{7}^{eff}>0$ and  $C_{7}^{eff}<0$ almost coincide with
each other, and the enhancement with respect to the SM is much more
sizable.
\item BR for  $B \rar \tau^+ \tau^- \gamma$ decay is at the order of
magnitude $10^{-9}$ ($10^{-7}$) in the SM and in model III without NHB
effects  for  $|r_{tb}|<1$ ( $r_{tb}>1$). However, including NHB exchanges
may enhance it almost two orders of magnitude compared to the SM
prediction, even for the smaller values of $\bar{\xi}_{N,\tau\tau}^{D}$.
\item $|A_{FB}|$ is at the order of magnitude
$10^{-1}$ ($10^{-2}$) for $ |r_{tb}| <1$ ($r_{tb} >1$) case and smaller
compared to the SM results, which is -0.181.
\item The 2HDM contributions change  $P_L$  and $<P_L>$ greatly compared
to the SM case and these quantities are very sensitive to the NHB effects.
\end{itemize}
In conclusion, we can say that  experimental investigation of BR, $A_{FB}$ 
and $P_L$ may provide an essential test for the effects of NHB exchanges
and new physics beyond the SM.
\newpage
\begin{appendix}
\section{The operator basis}
The operator basis in the  2HDM (model III ) for our process  
is \cite{Dai,Grinstein2,Misiak}
\begin{eqnarray}
 O_1 &=& (\bar{s}_{L \alpha} \gamma_\mu c_{L \beta})
               (\bar{c}_{L \beta} \gamma^\mu b_{L \alpha}), \nonumber   \\
 O_2 &=& (\bar{s}_{L \alpha} \gamma_\mu c_{L \alpha})
               (\bar{c}_{L \beta} \gamma^\mu b_{L \beta}),  \nonumber   \\
 O_3 &=& (\bar{s}_{L \alpha} \gamma_\mu b_{L \alpha})
               \sum_{q=u,d,s,c,b}
               (\bar{q}_{L \beta} \gamma^\mu q_{L \beta}),  \nonumber   \\
 O_4 &=& (\bar{s}_{L \alpha} \gamma_\mu b_{L \beta})
                \sum_{q=u,d,s,c,b}
               (\bar{q}_{L \beta} \gamma^\mu q_{L \alpha}),   \nonumber  \\
 O_5 &=& (\bar{s}_{L \alpha} \gamma_\mu b_{L \alpha})
               \sum_{q=u,d,s,c,b}
               (\bar{q}_{R \beta} \gamma^\mu q_{R \beta}),   \nonumber  \\
 O_6 &=& (\bar{s}_{L \alpha} \gamma_\mu b_{L \beta})
                \sum_{q=u,d,s,c,b}
               (\bar{q}_{R \beta} \gamma^\mu q_{R \alpha}),  \nonumber   \\  
 O_7 &=& \frac{e}{16 \pi^2}
          \bar{s}_{\alpha} \sigma_{\mu \nu} (m_b R + m_s L) b_{\alpha}
                {\cal{F}}^{\mu \nu},                             \nonumber  \\
 O_8 &=& \frac{g}{16 \pi^2}
    \bar{s}_{\alpha} T_{\alpha \beta}^a \sigma_{\mu \nu} (m_b R +
m_s L)  
          b_{\beta} {\cal{G}}^{a \mu \nu} \nonumber \,\, , \\  
 O_9 &=& \frac{e}{16 \pi^2}
          (\bar{s}_{L \alpha} \gamma_\mu b_{L \alpha})
              (\bar{\tau} \gamma^\mu \tau)  \,\, ,    \nonumber    \\
 O_{10} &=& \frac{e}{16 \pi^2}
          (\bar{s}_{L \alpha} \gamma_\mu b_{L \alpha})
              (\bar{\tau} \gamma^\mu \gamma_{5} \tau)  \,\, ,  \nnb \\
Q_1&=&   \frac{e^2}{16 \pi^2}(\bar{s}^{\alpha}_{L}\,b^{\alpha}_{R})\,(\bar{\tau}\tau )
\, , \nnb  \\ 
Q_2&=&\frac{e^2}{16 \pi^2}(\bar{s}^{\alpha}_{L}\,b^{\alpha}_{R})\,
(\bar{\tau} \gamma_5 \tau ) \, , \nnb \\
Q_3&=&    \frac{g^2}{16 \pi^2}(\bar{s}^{\alpha}_{L}\,b^{\alpha}_{R})\,
\sum_{q=u,d,s,c,b }(\bar{q}^{\beta}_{L} \, q^{\beta}_{R} ) \, ,\nnb \\
Q_4&=&  \frac{g^2}{16 \pi^2}(\bar{s}^{\alpha}_{L}\,b^{\alpha}_{R})\,
\sum_{q=u,d,s,c,b } (\bar{q}^{\beta}_{R} \, q^{\beta}_{L} ) \, , \nnb \\
Q_5&=&   \frac{g^2}{16 \pi^2}(\bar{s}^{\alpha}_{L}\,b^{\beta}_{R})\,
\sum_{q=u,d,s,c,b } (\bar{q}^{\beta}_{L} \, q^{\alpha}_{R} ) \, , \nnb \\
Q_6&=&   \frac{g^2}{16 \pi^2}(\bar{s}^{\alpha}_{L}\,b^{\beta}_{R})\,
\sum_{q=u,d,s,c,b } (\bar{q}^{\beta}_{R} \, q^{\alpha}_{L} ) \, , \nnb \\
Q_7&=&   \frac{g^2}{16 \pi^2}(\bar{s}^{\alpha}_{L}\,\sigma^{\mu \nu} \, 
b^{\alpha}_{R})\,
\sum_{q=u,d,s,c,b } (\bar{q}^{\beta}_{L} \, \sigma_{\mu \nu } 
q^{\beta}_{R} ) \, , \nnb \\
Q_8&=&    \frac{g^2}{16 \pi^2}(\bar{s}^{\alpha}_{L}\,\sigma^{\mu \nu} 
\, b^{\alpha}_{R})\,
\sum_{q=u,d,s,c,b } (\bar{q}^{\beta}_{R} \, \sigma_{\mu \nu } 
q^{\beta}_{L} ) \, ,  \nnb \\ 
Q_9&=&   \frac{g^2}{16 \pi^2}(\bar{s}^{\alpha}_{L}\,\sigma^{\mu \nu} 
\, b^{\beta}_{R})\,
\sum_{q=u,d,s,c,b }(\bar{q}^{\beta}_{L} \, \sigma_{\mu \nu } 
q^{\alpha}_{R} ) \, , \nnb \\
Q_{10}&= & \frac{g^2}{16 \pi^2}(\bar{s}^{\alpha}_{L}\,\sigma^{\mu \nu} \, 
b^{\beta}_{R})\,
\sum_{q=u,d,s,c,b }(\bar{q}^{\beta}_{R} \, \sigma_{\mu \nu } q^{\alpha}_{L} )
\label{op1}
\end{eqnarray}
where $\alpha$ and $\beta$ are $SU(3)$ colour indices and 
${\cal{F}}^{\mu \nu}$ and ${\cal{G}}^{\mu \nu}$ are the field strength 
tensors of the electromagnetic and strong interactions, respectively. Note 
that there are also flipped chirality partners of these operators, which 
can be obtained by interchanging $L$ and $R$ in the basis given above in 
model III. However, we do not present them here since corresponding  Wilson 
coefficients are negligible.
\section{The Initial values of the Wilson coefficients.}
The initial values of the Wilson coefficients for the relevant process 
in the SM are \cite{Grinstein2}
\begin{eqnarray}
C^{SM}_{1,3,\dots 6}(m_W)&=&0 \nonumber \, \, , \\
C^{SM}_2(m_W)&=&1 \nonumber \, \, , \\
C_7^{SM}(m_W)&=&\frac{3 x_t^3-2 x_t^2}{4(x_t-1)^4} \ln x_t+
\frac{-8 x_t^3-5 x_t^2+7 x_t}{24 (x_t-1)^3} \nonumber \, \, , \\
C_8^{SM}(m_W)&=&-\frac{3 x_t^2}{4(x_t-1)^4} \ln x_t+
\frac{-x_t^3+5 x_t^2+2 x_t}{8 (x_t-1)^3}\nonumber \, \, , \\ 
C_9^{SM}(m_W)&=&-\frac{1}{sin^2\theta_{W}} B(x_t) +
\frac{1-4 \sin^2 \theta_W}{\sin^2 \theta_W} C(x_t)-D(x_t)+
\frac{4}{9}, \nonumber \, \, , \\
C_{10}^{SM}(m_W)&=&\frac{1}{\sin^2\theta_W}
(B(x_t)-C(x_t))\nonumber \,\, , \\
C_{Q_i}^{SM}(m_W) & = & 0~~~ i=1,..,10
\end{eqnarray}
and for the additional part due to charged Higgs bosons are 
\begin{eqnarray}
C^{H}_{1,\dots 6 }(m_W)&=&0 \nonumber \, , \\
C_7^{H}(m_W)&=& Y^2 \, F_{1}(y_t)\, + \, X Y \,  F_{2}(y_t) 
\nonumber  \, \, , \\
C_8^{H}(m_W)&=& Y^2 \,  G_{1}(y_t) \, + \, X Y \, G_{2}(y_t) 
\nonumber\, \, , \\
C_9^{H}(m_W)&=&  Y^2 \,  H_{1}(y_t) \nonumber  \, \, , \\
C_{10}^{H}(m_W)&=& Y^2 \,  L_{1}(y_t)  
\label{CH} \, \, , 
\end{eqnarray}
where 
\bea
X & = & \frac{1}{m_{b}}~~~\left(\bar{\xi}^{D}_{N,bb}+\bar{\xi}^{D}_{N,sb}
\frac{V_{ts}}{V_{tb}} \right) ~~,~~ \nnb \\
Y & = & \frac{1}{m_{t}}~~~\left(\bar{\xi}^{U}_{N,tt}+\bar{\xi}^{U}_{N,tc}
\frac{V^{*}_{cs}}{V^{*}_{ts}} \right) ~~.~~
\eea
The NHB effects bring new operators and the corresponding Wilson
coefficients read as 
\bea
\!\!\!\!\!\!\!\!\!\!\!\!\!\!\!\!\!\!\!\!\!\!\!\!\!\!\!\!\!\!\!\!\!\!\!\!\!\!\!
\!\!\!\!\!\!\!\!\!\!\!\!\!\!\!\!\!\!\!\!\!\!\!\!\!\!\!\!\!\!\!\!\!\!\!\!\!\!\!
C^{A^{0}}_{Q_{2}}((\bar{\xi}^{U}_{N,tt})^{3}) =
\frac{\bar{\xi}^{D}_{N,\tau \tau}(\bar{\xi}^{U}_{N,tt})^{3}
m_{b} y_t (\Theta_5 (y_t)z_A-\Theta_1 (z_{A},y_t))}{32 \pi^{2}m_{A^{0}}^{2} 
m_{t} \Theta_1 (z_{A},y_t) \Theta_5 (y_t)} , \nnb
\eea
\bea
C^{A^{0}}_{Q_{2}}((\bar{\xi}^{U}_{N,tt})^{2})=\frac{\bar{\xi}^{D}_{N,\tau\tau}
(\bar{\xi}^{U}_{N,tt})^{2}
\bar{\xi}^{D}_{N,bb}}{32 \pi^{2}  m_{A^{0}}^{2}}\Big{(} 
\frac{(y_t (\Theta_1 (z_{A},y_t) - \Theta_5 (y_t) (xy+z_A))-
2 \Theta_1 (z_{A},y_t) \Theta_5 (y_t)   \ln [\frac{z_A \Theta_5 (y_t)}
{ \Theta_1 (z_{A},y_t)}]}{ \Theta_1 (z_{A},y_t) \Theta_5 (y_t)}\Big{)}, \nnb
\eea
\bea
C^{A^{0}}_{Q_{2}}(\bar{\xi}^{U}_{N,tt}) &=& 
\frac{g^2\bar{\xi}^{D}_{N,\tau\tau}\bar{\xi}^{U}_{N,tt} m_b
x_t}{64 \pi^2 m_{A^{0}}^{2}  m_t } \Bigg{(}\frac{2}{\Theta_5 (x_t)}
- \frac{xy x_t+2 z_A}{\Theta_1 (z_{A},x_t)}-2
\ln [\frac{z_A \Theta_5(x_t)}{ \Theta_1 (z_{A},x_t)}]\nnb \\ &
&\!\!\!\!\!\!\!\!\!\!\!\!\!\!\!\!\!\!\!\!\!\!\!- x y x_t y_t(\frac{(x-1) x_t 
(y_t/z_A-1)-(1+x)y_t)}{(\Theta_6 -(x-y)(x_t -y_t))(\Theta_3
(z_A)+(x-y)(x_t-y_t)z_A)}-
\frac{x (y_t+x_t(1-y_t/z_A))-2 y_t }{\Theta_6 \Theta_3 (z_A)}) \Bigg{)} \nnb
\eea
\bea
\!\!\!C^{A^{0}}_{Q_{2}}(\bar{\xi}^{D}_{N,bb}) =
\frac{g^2\bar{\xi}^{D}_{N,\tau\tau}\bar{\xi}^{D}_{N,bb}}{64 \pi^2 m^2_{A^{0}} }
\Big{(}1-
\frac{x^2_t y_t+2 y (x-1)x_t y_t-z_A (x^2_t+\Theta_6)}{ \Theta_3 (z_A)}+
\frac{x^2_t (1-y_t/z_A)}{\Theta_6}+2 \ln [\frac{z_A \Theta_6}{ \Theta_2 (z_{A},x)}]
\Big{)}\nnb 
\eea
\bea
\lefteqn{\!\!\!\!\!\!\!\!\!\!\!\!\!\!\!\!\!\!\!\!\!C^{H^{0}}_{Q_{1}}
((\bar{\xi}^{U}_{N,tt})^{2}) = 
\frac{g^2 (\bar{\xi}^{U}_{N,tt})^2 m_b m_{\tau}
}{64 \pi^2 m^2_{H^{0}} m^2_t } \Bigg{(}
\frac{x_t (1-2 y) y_t}{\Theta_5 (y_t)}+\frac{(-1+2 \cos^2 \theta_W) (-1+x+y) 
y_t} {\cos^2 \theta_W \Theta_4 (y_t)} } \nnb
\\ & &
+\frac{z_H (\Theta_1 (z_H,y_t) x y_t + 
\cos^2 \theta_W \,(-2 x^2 (-1+x_t) y y^2_t+x x_t y y^2_t-\Theta_8 z_H))}
{\cos^2 \theta_W \Theta_1 (z_H,y_t) \Theta_7 }\Bigg{)} ,             
\label{NHB}
\eea
\bea
\lefteqn{\!\!\!\!\!\!\!\!\!\!\!\!\!\!\!\!\!\!\!\!\!\!\!\!\!\!
\!\!\!\!\!\!\!\!\!\!\!\!\!\!\!\!\!\!\!\!\!C^{H^{0}}_{Q_{1}}
(\bar{\xi}^{U}_{N,tt}) 
=\frac{g^2 \bar{\xi}^{U}_{N,tt} \bar{\xi}^{D}_{N,bb} 
m_{\tau}}{64 \pi^2 m^2_{H^{0}} m_t } \Bigg{(}
\frac{(-1+2 \cos^2 \theta_W)\, y_t}{\cos^2 \theta_W \, \Theta_4 (y_t)}-
\frac{x_t y_t}{\Theta_5 (y_t)}+\frac {x_t y_t(x y-z_H)}
{\Theta_1 (z_H,y_t)} } 
\nnb
\\ & & 
+\frac{(-1+2 \cos^2 \theta_W) 
y_t z_H}{\cos^2\theta_W \Theta_7}-2 x_t\, \ln \Bigg{[}
\frac{\Theta_5 (y_t) z_H} {\Theta_1 (z_H,y_t)} \Bigg{]} \Bigg{)}  ,
\nnb
\eea
\bea
\lefteqn{ C^{H^0}_{Q_{1}}(g^4) =-\frac{g^4 m_b m_{\tau} x_t}
{128 \pi^2 m^2_{H^{0}} m^2_t} 
\Bigg{(} -1+\frac{(-1+2x) x_t}{\Theta_5 (x_t) + y (1-x_t)}+
\frac{2 x_t (-1+ (2+x_t) y)}{\Theta_5 (x_t)} } \nnb \\
& & 
-\frac{4 \cos^2 \theta_W (-1+x+y)+ x_t(x+y)} {\cos^2 \theta_W 
\Theta_4 (x_t)} +\frac{x_t (x (x_t (y-2 z_H)-4 z_H)+2 z_H)} {\Theta_1 
(z_H,x_t)} \nnb \\ 
& &
+\frac{y_t ( (-1+x) x_t z_H+\cos^2 \theta_W ( (3 x-y) z_H+x_t 
(2 y (x-1)- z_H (2-3 x -y))))}{\cos^2 \theta_W (\Theta_3 (z_H)+x 
(x_t-y_t) z_H)} \nnb
\\ & & 
+ 2\, ( x_t \ln \Bigg{[} \frac{\Theta_5 (x_t) z_H}{\Theta_1 (z_H,x_t)} 
\Bigg{]}+ \ln \Bigg{[} \frac{x(y_t-x_t) z_H-\Theta_3 (z_H)} {(\Theta_5 (x_t)+ 
y (1-x_t) y_t z_H} \Bigg{]} )\Bigg{)}  ,\nnb  
\eea
\bea
C^{h_0}_{Q_1}((\bar{\xi}^U_{N,tt})^3) &=&
-\frac{\bar{\xi}^D_{N,\tau\tau} (\bar{\xi}^U_{N,tt})^3 m_b y_t}
{32 \pi^2 m_{h^0}^2 m_t \Theta_1 (z_h,y_t) \Theta_5 (y_t)}
 \Big{(} \Theta_1 (z_h,y_t) (2 y-1) + \Theta_5 (y_t) (2 x-1) z_h \Big{)} \nnb 
\eea
\bea
C^{h_0}_{Q_1}((\bar{\xi}^U_{N,tt})^2)=
\frac{\bar{\xi}^D_{N,\tau\tau} (\bar{\xi}^U_{N,tt})^2 }
{32 \pi^2  m_{h^0}^2  } \Bigg{(}
\frac{ (\Theta_5 (y_t) z_h (y_t-1)(x+y-1)-\Theta_1 (z_h,y_t)( \Theta_5(y_t)+y_t )
}{\Theta_1 (z_h)\Theta_5(y_t)}- 2 \ln \Bigg{[} \frac{z_h \Theta_5
(y_t)}{\Theta_1 (z_h)} \Bigg{]} \Bigg{)}\nnb 
\eea
\bea
C^{h^0}_{Q_{1}}(\bar{\xi}^{U}_{N,tt}) & = & -\frac{g^2
\bar{\xi}^{D}_{N,\tau\tau}\bar{\xi}^{U}_{N,tt} m_b x_t}{64 \pi^2 m^2_{h^{0}} 
m_t} \Bigg{(}\frac{2 (-1+(2+x_t) y)}{\Theta_5 (x_t)}-\frac{x_t
(x-1)(y_t-z_h)}{\Theta'_2 (z_h)}+2 \ln \Bigg{[}\frac{z_h \Theta_5
(x_t)}{\Theta_1 (z_h,x_t)} \Bigg{]} \nnb \\ & + & \frac{x (x_t (y-2 z_h)-
4 z_h)+2 z_h}{\Theta_1 (z_h,x_t)}  -  \frac{(1+x) y_t z_h}{x y x_t y_t+z_h
((x-y)(x_t-y_t)- \Theta_6)} \nnb \\ 
& + & 
\frac{\Theta_9 + y_t z_h ( (x-y)(x_t-y_t)-\Theta_6 )(2
x-1)}{z_h \Theta_6 (\Theta_6 -(x-y)(x_t-y_t))}+\frac{x (y_t z_h + x_t
(z_h-y_t))-2 y_t z_h}{\Theta_2 (z_h)} \Bigg{)}, \nnb
\eea
\bea
C^{h^0}_{Q_{1}}(\bar{\xi}^{D}_{N,bb}) &  = & -\frac{g^2
\bar{\xi}^{D}_{N,\tau\tau}\bar{\xi}^{D}_{N,bb}}{64 \pi^2 m^2_{h^0} }
\Bigg{(}\frac{y x_t y_t (x x^2_t(y_t-z_h)+\Theta_6 z_h
(x-2))}{z_h\Theta_2 (z_h)\Theta_6 }+2 \ln \Bigg{[}\frac{\Theta_6}{x_t y_t} \Bigg{]}
+2 \ln \Bigg{[}\frac{x_t y_t z_h}{\Theta_2 (z_h)} \Bigg{]}
\Bigg{)} \nnb 
\eea
where 
\bea
\Theta_1 (\omega , \lambda ) & = & -(-1+y-y \lambda ) \omega -x (y \lambda
+\omega - \omega \lambda ) \nnb \\
\Theta_2 (\omega ) & = &  (x_t +y (1-x_t)) y_t \omega - x x_t (y
y_t+(y_t-1) \omega)   \nnb \\
\Theta^{\prime}_2  (\omega ) & = & \Theta_2 (\omega , x_t \leftrightarrow y_t)    \nnb \\
\Theta_3 (\omega) & = & (x_t (-1+y)-y ) y_t \omega +
x x_t (y y_t+\omega(-1+y_t)) \nnb \\
\Theta_4 (\omega) & = & 1-x +x  \omega  \nnb \\
\Theta_5 (\lambda) & = & x + \lambda (1-x) \nnb \\
\Theta_6  & = & (x_t +y  (1-x_t))y_t +x x_t  (1-y_t) \nnb \\
\Theta_7  & = & (y (y_t -1)-y_t) z_H+x (y y_t + (y_t-1) z_H ) \\ 
\Theta_8  & = & y_t (2 x^2 (1+x_t) (y_t-1) +x_t (y(1-y_t)+y_t)+x
(2(1-y+y_t) \nnb \\ & + & x_t (1-2 y (1-y_t)-3 y_t))) \nnb \\
\Theta_9  & = & -x^2_t (-1+x+y)(-y_t+x (2 y_t-1)) (y_t-z_h)-x_t y_t z_h
(x(1+2 x)-2 y) \nnb \\ & + & y^2_t (x_t (x^2 -y (1-x))+(1+x) (x-y) z_h) 
\nnb
\eea
and
\begin{eqnarray}
& & x_t=\frac{m_t^2}{m_W^2}~~~,~~~y_t=
\frac{m_t^2}{m_{H^{\pm}}}~~~,~~~z_H=\frac{m_t^2}{m^2_{H^0}}~~~,~~~
z_h=\frac{m_t^2}{m^2_{h^0}}~~~,~~~ z_A=\frac{m_t^2}{m^2_{A^0}}~~~,~~~ \nnb
\end{eqnarray}
The explicit forms of the functions $F_{1(2)}(y_t)$, $G_{1(2)}(y_t)$, 
$H_{1}(y_t)$ and $L_{1}(y_t)$ in eq.(\ref{CH}) are given as
\begin{eqnarray}
F_{1}(y_t)&=& \frac{y_t(7-5 y_t-8 y_t^2)}{72 (y_t-1)^3}+
\frac{y_t^2 (3 y_t-2)}{12(y_t-1)^4} \,\ln y_t \nonumber  \,\, , 
\\ 
F_{2}(y_t)&=& \frac{y_t(5 y_t-3)}{12 (y_t-1)^2}+
\frac{y_t(-3 y_t+2)}{6(y_t-1)^3}\, \ln y_t 
\nonumber  \,\, ,
\\ 
G_{1}(y_t)&=& \frac{y_t(-y_t^2+5 y_t+2)}{24 (y_t-1)^3}+
\frac{-y_t^2} {4(y_t-1)^4} \, \ln y_t
\nonumber  \,\, ,
\\ 
G_{2}(y_t)&=& \frac{y_t(y_t-3)}{4 (y_t-1)^2}+\frac{y_t} {2(y_t-1)^3} \, 
\ln y_t  \nonumber\,\, ,
\\
H_{1}(y_t)&=& \frac{1-4 sin^2\theta_W}{sin^2\theta_W}\,\, \frac{xy_t}{8}\,
\left[ \frac{1}{y_t-1}-\frac{1}{(y_t-1)^2} \ln y_t \right]\nonumber \\
&-&
y_t \left[\frac{47 y_t^2-79 y_t+38}{108 (y_t-1)^3}-
\frac{3 y_t^3-6 y_t+4}{18(y_t-1)^4} \ln y_t \right] 
\nonumber  \,\, , 
\\ 
L_{1}(y_t)&=& \frac{1}{sin^2\theta_W} \,\,\frac{x y_t}{8}\, 
\left[-\frac{1}{y_t-1}+ \frac{1}{(y_t-1)^2} \ln y_t \right]
\nonumber  \,\, .
\\ 
\label{F1G1}
\end{eqnarray}
Finally, the initial values of the coefficients in the model III are
\begin {eqnarray}   
C_i^{2HDM}(m_{W})&=&C_i^{SM}(m_{W})+C_i^{H}(m_{W}) , \nnb \\
C_{Q_{1}}^{2HDM}(m_{W})&=& \int^{1}_{0}dx \int^{1-x}_{0} dy \,
(C^{H^{0}}_{Q_{1}}((\bar{\xi}^{U}_{N,tt})^{2})+
 C^{H^{0}}_{Q_{1}}(\bar{\xi}^{U}_{N,tt})+
 C^{H^{0}}_{Q_{1}}(g^{4})+C^{h^{0}}_{Q_{1}}((\bar{\xi}^{U}_{N,tt})^{3}) \nnb
\\ & + &
 C^{h^{0}}_{Q_{1}}((\bar{\xi}^{U}_{N,tt})^{2})+
 C^{h^{0}}_{Q_{1}}(\bar{\xi}^{U}_{N,tt})+
 C^{h^{0}}_{Q_{1}}(\bar{\xi}^{D}_{N,bb})) , \nnb  \\
 C_{Q_{2}}^{2HDM}(m_{W})&=& \int^{1}_{0}dx \int^{1-x}_{0} dy\,
(C^{A^{0}}_{Q_{2}}((\bar{\xi}^{U}_{N,tt})^{3})+
C^{A^{0}}_{Q_{2}}((\bar{\xi}^{U}_{N,tt})^{2})+
 C^{A^{0}}_{Q_{2}}(\bar{\xi}^{U}_{N,tt})+
 C^{A^{0}}_{Q_{2}}(\bar{\xi}^{D}_{N,bb}))\nnb \\
C_{Q_{3}}^{2HDM}(m_W) & = & \frac{m_b}{m_{\tau} \sin^2 \theta_W} 
 (C_{Q_{1}}^{2HDM}(m_W)+C_{Q_{2}}^{2HDM}(m_W)) \nnb \\
C_{Q_{4}}^{2HDM}(m_W) & = & \frac{m_b}{m_{\tau} \sin^2 \theta_W} 
 (C_{Q_{1}}^{2HDM}(m_W)-C_{Q_{2}}^{2HDM}(m_W)) \nnb \\
C_{Q_{i}}^{2HDM} (m_W) & = & 0\,\, , \,\, i=5,..., 10.
\label{CiW}
\end{eqnarray}
Here, we present $C_{Q_{1}}$ and $C_{Q_{2}}$ in terms of the Feynmann
parameters $x$ and $y$ since the integrated results are extremely large.
Using these initial values, we can calculate the coefficients 
$C_{i}^{2HDM}(\mu)$ and $C^{2HDM}_{Q_i}(\mu)$ 
at any lower scale in the effective theory 
with five quarks, namely $u,c,d,s,b$ similar to the SM case 
\cite{Misiak,Chao,Alil2,Buras}. 

The Wilson  coefficients playing  the essential role 
in this process are $C_{7}^{2HDM}(\mu)$, $C_{9}^{2HDM}(\mu)$,
$C_{10}^{2HDM}(\mu)$, 
$C^{2HDM}_{Q_1}(\mu )$ and $C^{2HDM}_{Q_2}(\mu )$. For completeness,
in the following we give their explicit expressions. 
\begin{eqnarray}
C_{7}^{eff}(\mu)&=&C_{7}^{2HDM}(\mu)+ Q_d \, 
(C_{5}^{2HDM}(\mu) + N_c \, C_{6}^{2HDM}(\mu))\nonumber \, \, ,
\label{C7eff}
\end{eqnarray}
where the LO  QCD corrected Wilson coefficient 
$C_{7}^{LO, 2HDM}(\mu)$ is given by
\begin{eqnarray} 
C_{7}^{LO, 2HDM}(\mu)&=& \eta^{16/23} C_{7}^{2HDM}(m_{W})+(8/3) 
(\eta^{14/23}-\eta^{16/23}) C_{8}^{2HDM}(m_{W})\nonumber \,\, \\
&+& C_{2}^{2HDM}(m_{W}) \sum_{i=1}^{8} h_{i} \eta^{a_{i}} \,\, , 
\label{LOwils}
\end{eqnarray}
and $\eta =\alpha_{s}(m_{W})/\alpha_{s}(\mu)$, $h_{i}$ and $a_{i}$ are 
the numbers which appear during the evaluation \cite{Buras}. 

$C_9^{eff}(\mu)$ contains a perturbative part and a part coming from LD
effects due to conversion of the real $\bar{c}c$ into lepton pair $\tau^+
\tau^-$:
\begin{eqnarray}
C_9^{eff}(\mu)=C_9^{pert}(\mu)+ Y_{reson}(s)\,\, ,
\label{C9efftot}
\end{eqnarray}
where
\begin{eqnarray} 
C_9^{pert}(\mu)&=& C_9^{2HDM}(\mu) \nonumber 
\\ &+& h(z,  s) \left( 3 C_1(\mu) + C_2(\mu) + 3 C_3(\mu) + 
C_4(\mu) + 3 C_5(\mu) + C_6(\mu) \right) \nonumber \\
&- & \frac{1}{2} h(1, s) \left( 4 C_3(\mu) + 4 C_4(\mu) + 3
C_5(\mu) + C_6(\mu) \right) \\
&- &  \frac{1}{2} h(0,  s) \left( C_3(\mu) + 3 C_4(\mu) \right) +
\frac{2}{9} \left( 3 C_3(\mu) + C_4(\mu) + 3 C_5(\mu) + C_6(\mu)
\right) \nonumber \,\, ,
\label{C9eff2}
\end{eqnarray}
and
\begin{eqnarray}
Y_{reson}(s)&=&-\frac{3}{\alpha^2_{em}}\kappa \sum_{V_i=\psi_i}
\frac{\pi \Gamma(V_i\rightarrow \tau^+ \tau^-)m_{V_i}}{q^2-m_{V_i}+i m_{V_i}
\Gamma_{V_i}} \nonumber \\
& & \left( 3 C_1(\mu) + C_2(\mu) + 3 C_3(\mu) + 
C_4(\mu) + 3 C_5(\mu) + C_6(\mu) \right).
\label{Yres}
\end{eqnarray}
In eq.(\ref{C9efftot}), the functions $h(u, s)$ are given by
\begin{eqnarray}
h(u, s) &=& -\frac{8}{9}\ln\frac{m_b}{\mu} - \frac{8}{9}\ln u +
\frac{8}{27} + \frac{4}{9} x \\
& & - \frac{2}{9} (2+x) |1-x|^{1/2} \left\{\begin{array}{ll}
\left( \ln\left| \frac{\sqrt{1-x} + 1}{\sqrt{1-x} - 1}\right| - 
i\pi \right), &\mbox{for } x \equiv \frac{4u^2}{ s} < 1 \nonumber \\
2 \arctan \frac{1}{\sqrt{x-1}}, & \mbox{for } x \equiv \frac
{4u^2}{ s} > 1,
\end{array}
\right. \\
h(0,s) &=& \frac{8}{27} -\frac{8}{9} \ln\frac{m_b}{\mu} - 
\frac{4}{9} \ln s + \frac{4}{9} i\pi \,\, , 
\label{hfunc}
\end{eqnarray}
with $u=\frac{m_c}{m_b}$.
The phenomenological parameter $\kappa$ in eq. (\ref{Yres}) is taken as 
$2.3$. In Eqs. (37) and (\ref{Yres}), the contributions of 
the coefficients $C_1(\mu)$, ...., $C_6(\mu)$ are due to the operator mixing.

Finally, the Wilson coefficients $C_{Q_1}(\mu)$ and $C_{Q_2}(\mu )$  
are given by \cite{Dai}
\beq
C_{Q_i}(\mu )=\eta^{-12/23}\,C_{Q_i}(m_W)~,~i=1,2~. 
\eeq
\end{appendix}
\newpage
\newpage
\end{document}